\documentclass[usenatbib]{mnras}
\usepackage{amsmath}
\usepackage{graphicx,times}
\usepackage{txfonts}

\title[Merging galaxy clusters with very hot shock fronts]{Two merging galaxy clusters with very hot shock fronts observed shortly before pericentric passage}

\author[Lagan\'{a}, Souza, Machado et al.]{T.~F.~Lagan\'{a},$^{1}$\thanks{E-mail: tatiana.lagana@cruzeirodosul.edu.br}
G.~S.~Souza,$^{1}$
R.~E.~G.~Machado,$^{2}$
R.~C.~Volert$^{2}$ and
P.~A.~A.~Lopes$^{3}$\\
$^{1}$N\'ucleo de Astrof\'\i sica, Universidade Cruzeiro do Sul / Universidade Cidade de S\~ao Paulo, Rua Galv\~{a}o Bueno, 868, 01506-000, S\~{a}o Paulo SP, Brazil\\
$^{2}$Departamento Acad\^emico de F\'isica, Universidade Tecnol\'ogica Federal do Paran\'a, Rua Sete de Setembro 3165, 80230-901 Curitiba PR, Brazil\\
$^{3}$Observat\'{o}rio do Valongo, Universidade Federal do Rio de Janeiro, Ladeira do Pedro Ant\^{o}nio 43, 20080-090, Rio de Janeiro RJ, Brazil\\
}

\date{Accepted 2019 June 3. Received 2019 May 21; in original form 2019 April 12.}
\pubyear{2019}

\begin{document}  
\label{firstpage}
\pagerange{\pageref{firstpage}--\pageref{lastpage}}
\maketitle

\begin{abstract}
We present a detailed analysis of two merging clusters, from XMM-\textit{Newton} X-ray archival data: PLCKESZ G036.7+14.9 ($z=0.15$; hereafter G036) and PLCK G292.5+22.0 ($z=0.30$; hereafter G292). We notice that the intracluster medium is heated as a result of the merger, and we find evidence for a merger shock in the region between both subcluster haloes. X-ray observations confirm that the shocks in these systems are among the hottest known in the literature.
From the ICM analysis of temperature discontinuity, the Mach numbers were determined to be $M_{\rm G036}=1.3$ and $M_{\rm G292}=1.5$ for G036 and G292, respectively. In this paper, for each cluster, we propose a hydrodynamic model for the merger as a whole, compatible with their diffuse X-ray emission and temperature maps. The simulations suggest that both clusters are observed shortly before pericentric passage. Our simulation results indicate that the merger of the G036 system is seen at an inclination of 50$^{\circ}$ (the angle between the plane of the orbit and the plane of the sky), and merely 50 Myr prior to the pericentric passage. In the case of G292, the subclusters would be merging not far from the plane of the sky ($i=18^{\circ}$) and are observed 150\,Myr before the two cores collide.
\end{abstract}

\begin{keywords}
X-ray: galaxies: clusters -- galaxies: clusters: individual (PLCK G036.7+14.9) -- galaxies: clusters: individual (PLCK G292.5+22.0) -- galaxies: clusters: intracluster medium -- methods: numerical
\end{keywords}

\section{Introduction}

Clusters of galaxies are the largest gravitationally bound structures in the Universe, that according to the hierarchical  structure  formation  scenario  are formed through  a  series  of  successive  mergers  of  smaller  clusters and  groups  of  galaxies,  as  well  as  through  continuous  
accretion of surrounding matter over cosmic time.
Cluster  mergers  are  very  energetic events,  releasing  energies  up  to 10$^{64} \rm ~erg\,s^{-1}$  on  a  few Gyr timescale. 
This energy is dissipated through low-Mach number shocks and turbulence \citep[e.g.,][]{Markevitch07}, heating the  intracluster medium (ICM).

%ESTUDOS DE AGLOMERADOS. IMPORTANCIA
%which are formed through the 
%mergers of smaller structures over cosmic time. 

During merger processes, turbulence and shock waves are able to accelerate charged relativistic particles \citep{DeYoung92} that in magnetized regions of clusters can be revealed through
radio observations of the non-thermal synchrotron emission. Such phenomena are
known as radio relics and halos \citep[for reviews see][]{Feretti2012,Brunetti2014}.
 %Merger  events  can  be  revealed  in  the  radio band, via non-thermal synchrotron emission from diffuse sources not directly related to cluster galaxies.  Indeed, part  of  the  energy  released  by  a  cluster  merger  maybe used to amplify the magnetic field and to accelerate relativistic particles.  Results of such phenomena are the so called radio relics and halos, depending on their position in the cluster and on their morphological, spectral and polarization properties \citep[for reviews see][]{Feretti2012,Brunetti2014}. 
Also, X-ray observations show evidence of shocked gas in merging galaxy clusters that are now relatively common \citep{Fabian03,Markevitch04,Markevitch07,Canning17,Emery17}. 
Shocks have been commonly observed from X-ray data from XMM-\textit{Newton} \citep[e.g.][]{Ogrean2013}, \textit{Chandra} \citep[e.g.][]{Russell2010} and \textit{Suzaku} \citep[e.g.][]{Akamatsu2012}.
In particular, X-ray surface brightness and temperature maps show features that strongly suggest they result from shocks driven by the supersonic motion of merging subclusters. A shock in the ICM will result in both a density and temperature discontinuity in the gas, which creates an X-ray excess and temperature jump.  Using X-ray spectroscopy to measure the 
temperature and density of the gas on either side of the shock, together 
with the  Rankine-Hugoniot jump conditions, can provide a
reliable estimate of the shock  velocity. Thus, X-ray observations of merging shocks currently provide the only method for determining the velocity of the cluster gas in the plane of the sky \citep[e.g.,][]{MSV99}, and are a  key observational tool in the study of these systems.

%\textbf{Previous studies of merging galaxy clusters have demonstrated that physical process can be separated to three topics: early (or pre), ongoing and late (or post) stage cluster mergers. There are studies of ongoing and late mergers in the literature \citep[e.g.,][]{Briel91,Markevitch2002,Kempner04, Russell10}. However, early stages of mergers are still quite a few \citep{Fujita96, Akamatsu16,Bulbul16}. E DAI.... ESTA SOLTO ISSO}

In a complementary approach, merging clusters have frequently been studied by means of hydrodynamical simulations. This technique has proven to be fruitful in investigating merger events in general, in order to explore from a theoretical standpoint the physical mechanisms that shape the intracluster medium (ICM) \citep[e.g.][]{ZuHone2010, ZuHone2011, ZuHone2013a, Vazza2012, Iapichino2017, Schmidt2017}. But simulations of binary collisions are particularly useful to model specific observed clusters \citep[e.g.][]{Springel2007, Mastropietro2008, vanWeeren2011, Machado2013, Lage2014, Donnert2014, Molnar2015, Machado2015, Walker2018, Halbesma2019}. The results of such simulations often provide insight into the probable history of a given observed cluster, thus aiding the interpretation of its current dynamical state. Under certain circumstances, triple \citep{Bruggen2012} and quadruple \citep{Ruggiero2019} mergers may be modeled by simulations.

Shock fronts are a particular feature of interest that are well suited to be studied via simulations \citep[e.g.][]{Springel2007, Machado2013}. The typical collision velocities of clusters are such that they drive supersonic shock waves with expected Mach numbers of $\mathcal{M}\lesssim3$ \citep{Sarazin2002}. 
Mach numbers derived from observations are typically in the range $\mathcal{M} \sim 2-3$. Such estimates usually suffer from limited photon counts and from the unknown inclination of the merger. In fact, \cite{Hong2014} points out that in the situations where shocks are detected from X-ray and also from radio data, the estimated shock velocities tend to disagree. In statistical analyses of cosmological simulations \cite[e.g.][]{Vazza2009, Vazza2011}, strong shocks are not rare. For instance, \cite{Planelles2013} obtain $\mathcal{M}\sim5$ for the mean mass-weighted Mach number from hydrodynamical cosmological simulations. From the Illustris simulation \citep{Vogelsberger2014,Genel2014}, \cite{Schaal2015} find that nearly 75 per cent of shocks have Mach numbers $\mathcal{M}<6$, regardless of redshift. Depending on the time scale, strong simulated shock fronts may not be detectable in X-ray data, presumably because they have already travelled to the low-density outskirts of the clusters \citep{MachadoMonteiro2015, Monteiro-Oliveira2017}. So shock fronts are more likely to be detectable shortly before or shortly after central passage.

%Consequently, the physical mechanisms of early stage mergers are still poorly understood, and observations of new objects are critical to understand the mechanism.

Here, we present an investigation of two merging galaxy clusters based on X-ray analysis and numerical simulation results that put constraints on the stage of the merger. We offer simulated models that aim to reproduce some of the main features of the merging clusters studied here, namely: their masses, the separation between them, the shock and pre-shock temperatures, and to a lesser degree the morphology of the shock fronts. The paper is structured as follows. In Section \ref{sample} we describe the two merging systems analysed in this work as well as the XMM-\textit{Newton} data reduction and the numerical simulation setup. Then, in Section \ref{res} we present our results and proceed with the discussion. We conclude our findings in Section \ref{conc}.

In this paper, we assume a standard $\Lambda$ cold dark matter ($\Lambda$CDM) cosmology, with $H_{0} = 73\,{\rm km\,s^{-1}\,Mpc^{-1}}$, the matter density parameter $\Omega_{\rm M} = 0.3$, dark energy density parameter $\Omega_{\Lambda} =0.7$, so that at the redshift of G036 ($z=0.15$) 1 arcsec corresponds to 2.614\,kpc and at the redshift of G292 ($z=0.3$) 1 arcsec corresponds to 4.454\,kpc. Errors quoted in the spectral analysis are in the 68\% confidence limit.

%%%%%%%%%%%%%%%%%%%%%%%%%%%%%%%%%%%%%%%%%%%%%%%%%%%
\section{Sample and Methods}
\label{sample}

\subsection{Observational Data}

In this study we present XMM-\textit{Newton} analysis and numerical simulation results of two merging clusters: PLCK G292.5+22.0 (G292, hereafter) and PLCK G036.7+14.9 (G036, hereafter).

\begin{figure*}
\centering
\includegraphics[scale=0.5]{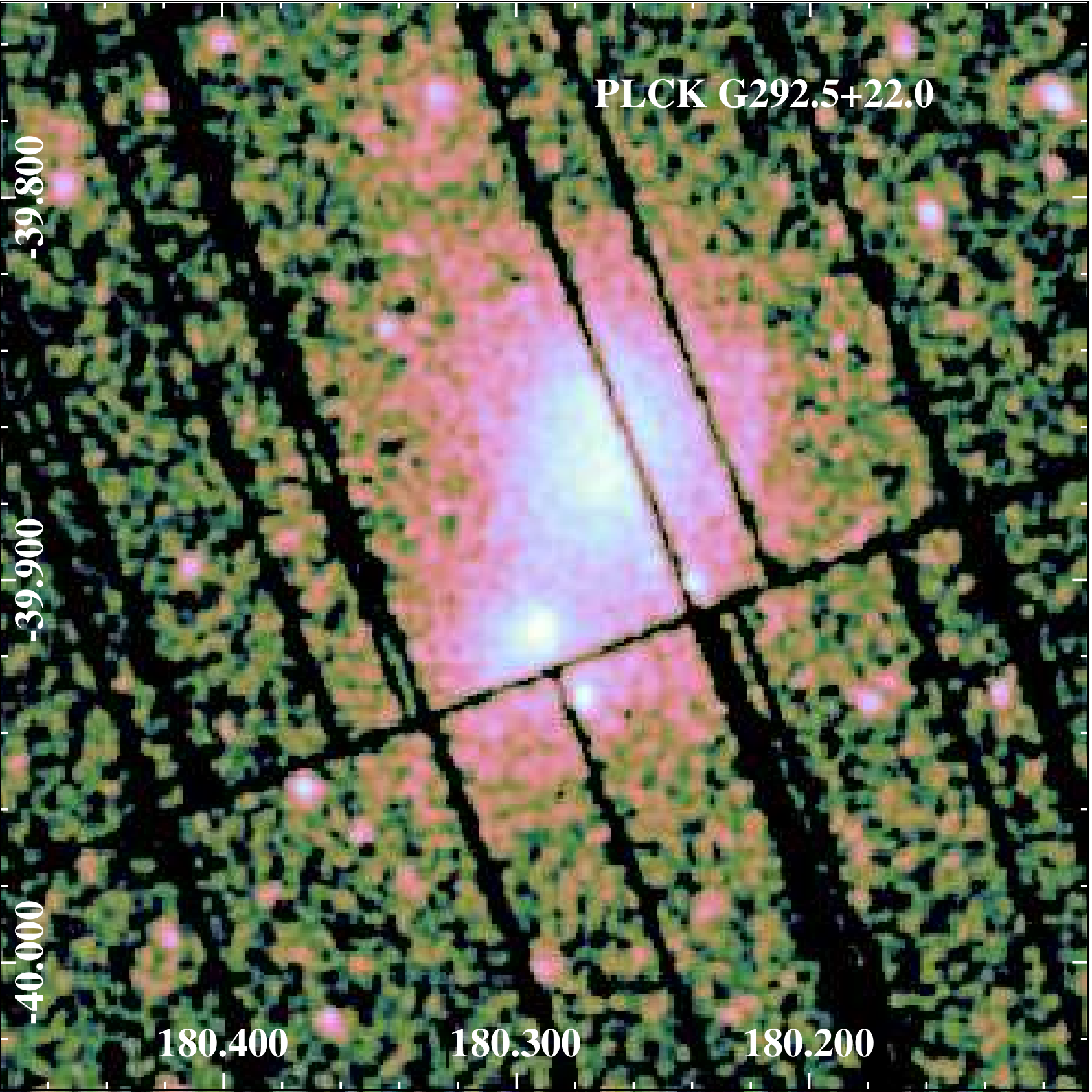}\quad
\includegraphics[scale=0.5]{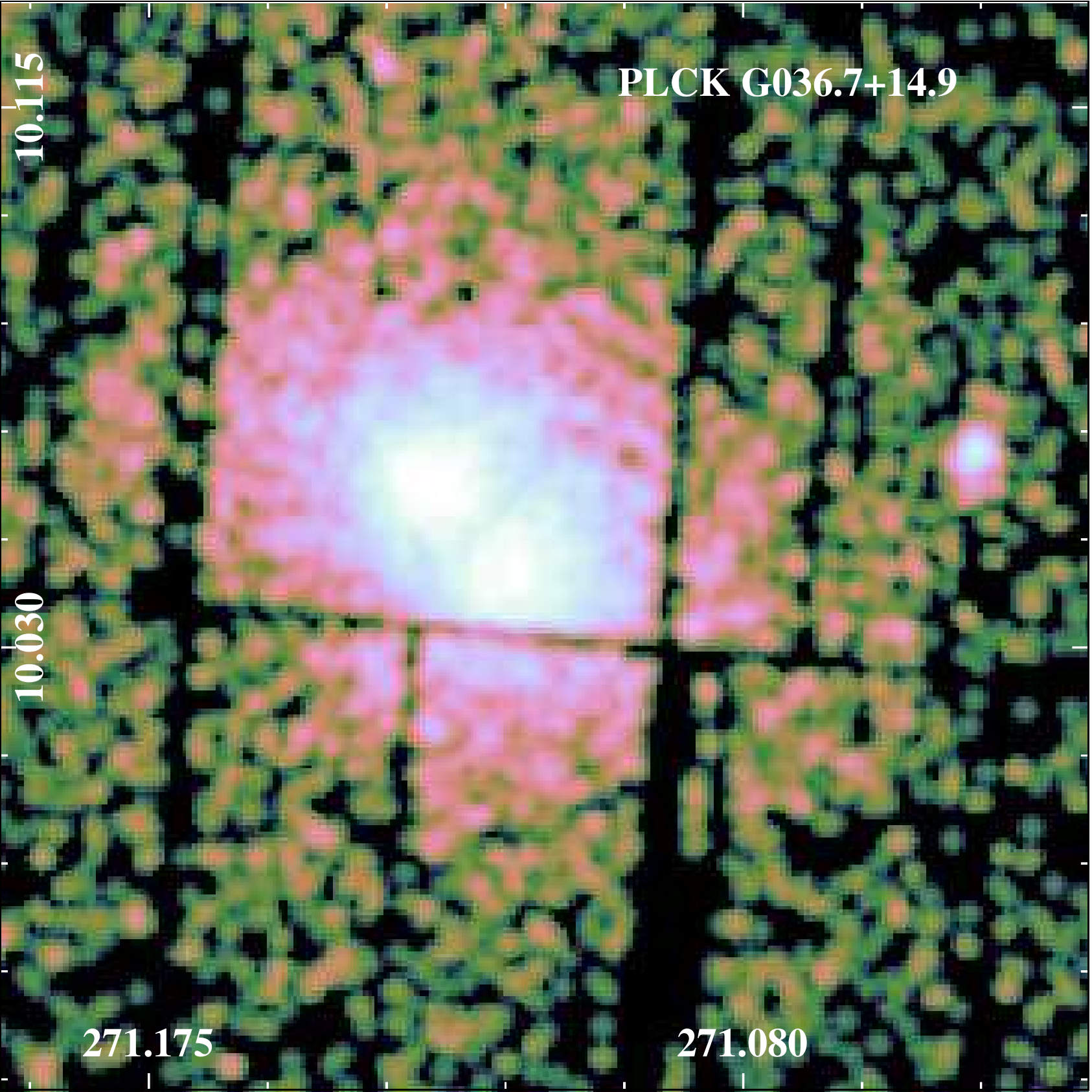}\quad
\caption{X-ray images for G292 (left panel) and G036 (right panel). North is up and west to the right.}
\label{fig:imgrx}
\end{figure*}

G292  was observed by XMM-\textit{Newton} (ObsID 0674380701) on 2011 December 11th for  almost 58 ks. Prime full frame mode was used for the three cameras with medium filters. The XMM-\textit{Newton} image (Fig.~\ref{fig:imgrx}, left panel)  shows that large scale emission for G292 is elongated along the northwest-southeast direction. Thus, in the left panel of Fig.~\ref{fig:imgrx}, the sub-cluster to the northwest (towards the upper right side of the frame) is called G292N (R.A.= 12:01:04.644 and DEC = $-$39:51:47.32). The other sub-cluster (towards the lower left side) is G292S (R.A. = 12:01:10.483 and DEC = $-$39:54:46.52). 

G036 (ObsID 0692931901) was observed on 2013 March 3rd for around 15 ks. Prime full frame mode was used for the three cameras with medium filters  for MOS1 and MOS2 and thin filters for pn. For G036 (Fig.~\ref{fig:imgrx}, right panel), the X-ray  emission shows a northeast-southwest elongation  with two sub-clusters  close in projection, but still separated. In the right panel of Fig.~\ref{fig:imgrx}, the sub-cluster to the northeast (towards the upper left of the frame) is called G036N, centered at R. A. = 18:04:30.847 and DEC =  +10:03:17.97) and the other subsystem (towards the lower right of the frame) is G036S (R.A. = 18:04:27.813 and DEC = +10:02:29.97). The basic properties of these clusters are presented in Table~\ref{tab:obs}.

%We performed a detailed imaging and spectral analysis that  revealed two pairs of clusters undergoing major mergers that were confirmed by our hydrodynamical  
%simulation results. 

%--------------------------------------------------
\begin{table*}
\centering
\caption{XMM ObsID, coordinates, redshifts and hydrogen column densities in the directions of each cluster.}

\label{tab:obs}
\begin{tabular}{cccccc}
\hline
Cluster & ObsID & R.A. & DEC & $z$ &   $n_{\rm H}$ \\
(Plank Name) & & (J2000) & (J2000) &   & $10^{21} {\rm cm^{-2}}$\\
\hline
PLCK G292.5+22.0 & 0674380701 &12:01:08.18 & -39:54:33.0 & 0.30 &1.19\\
PLCK G036.7+14.9 & 0692931901 &18:04:29.70 & ~10:02:43.7 & 0.15 &1.34 \\
\hline
\end{tabular}
\end{table*}
%--------------------------------------------------

%--------------------------------------------------
%--------------------------------------------------

%%%%%%%%%%%%%%%%%%%%%%%%%%%%%%%%%%%%%%%%%%%%%%%%%%%
\subsection{XMM-Newton Data Reduction}
\label{sec:XMM}

Data reduction was done with SAS version 6.8.0 and calibration files updated to 2016 July.  In order to filter background flares  we applied a 2$\sigma$-clipping procedure using the light curves in the [1--10]\,keV energy band. To take into account each detector background contribution, we obtained a background spectrum in an outer annulus of the observation, and in the 10--12 keV energy band. We compared these spectra with the one obtained by \citet{ReadPonman03} blank sky in the same detector region and energy band. Then, we rescaled the observation background to the blank sky background to obtain a normalisation parameter that will be used in the spectral fits. Point sources were detected by visual inspection, confirmed in the High Energy Catalogue 2XMMi Source, and excluded from our analysis. G036N hosts an unresolved radio source in its center \citep[NVSS J180431$+$100323,][]{Zhang15} that was excluded from our analysis.

The spectral analysis was restricted to the energy range  [0.7--7.0]\,keV and grouped to contain a minimum of 9 counts per spectral channel. We also excluded the energy band from 1.2 to 1.9 keV to avoid any influence from Al and Si instrumental lines. A single temperature fit was adopted to model the cluster spectrum using an absorbed thermal plasma emission model WABS(MEKAL) \citep{Balucinska92,Morrison83,KM93} in XSPEC version 12.9.1. In the direction of the clusters, the molecular hydrogen column density was not negligible. Thus the hydrogen column density ($N_{H}$) was considered as the sum of the weighted average atomic hydrogen column density listed in the Leiden/Argentine/Bonn (LAB) and the molecular hydrogen column density. For all fits, $N_{H}$ was fixed at the values listed in Table~\ref{tab:obs}, letting all other parameters (e.g.~temperature, metallicity and normalization) vary. Abundances were measured assuming the ratios from \citet{Asplund09}.

%%%%%%%%%%%%%%%%%%%%%%%%%%%%%%%%%%%%%%%%%%%%%%%%%%%
\subsubsection{2D Spectral Maps}

We have derived 2D temperature ($kT$), pseudo-entropy ($S$) and pseudo-pressure ($P$) maps for these  merging clusters. 
To construct 2D maps we subdivided the data into small pixels in which we extracted spectra. we imposed a minimum count number of 900 (after background subtraction, corresponding to a S/N of at least 30), necessary for obtaining a good spectral fit. This is done for all the pixels in the grid, and the spectra of the three EPIC instruments (MOS1, MOS2 and pn) are then simultaneously fitted. This procedure \citep[already described in][]{Durret10, Durret11, Lagana15, Lagana19} allows us to perform a reliable spectral analysis in each spectral bin, in order to derive $kT$, $S$ and $P$ maps.

%Each pixel  is 512 $\times$ 512 XMM-{\it Newton}  EPIC physical pixels (that is 25.6 arcsec$^{2}$) and we imposed a minimum count number of 900 (after background subtraction, corresponding to a S/N of at least 30), necessary for obtaining a good spectral fit. We can increase the area of this small region 
%up to a region of 5 $\times$ 5 pixel to obtain the fit. If we still do not have enough counts, this region is ignored and we proceed to the next neighbouring pixel. This is done for all the pixels in the grid, and the spectra of the three EPIC instruments (MOS1, MOS2 and pn) are then simultaneously fitted. This procedure \citep[already described in][]{Durret10, Durret11, Lagana15, Lagana19} allows us to perform a reliable spectral analysis in each spectral bin, in order to derive $kT$, $S$ and $P$ maps.}

%%%%%%%%%%%%%%%%%%%%%%%%%%%%%%%%%%%%%%%%%%%%%%%%%%%
\subsection{Numerical Simulations}

% goal
Hydrodynamical $N$-body simulations were performed with the goal of obtaining suitable models that represent the merging events of both clusters. Our main constraints were the masses of the clusters, their separations and the temperature jumps of the shock fronts. Aiming to satisfy these criteria, we started from educated guesses and ran several simulations by trial and error until satisfactory models were reached. Here we present the methods used to create the initial conditions and to perform the simulations.

% ic
The simulations are idealized binary mergers, where two initially spherical and relaxed clusters collide with a given initial velocity and a given impact parameter. Each cluster is represented by dark matter particles and gas particles. In our models, the dark matter halo follows a \cite{Hernquist1990} profile, which is similar to an NFW \citep*{NFW1997} profile in the inner parts. For the gas, a \cite{Dehnen1993} profile is adopted, with such a choice of parameters that it resembles a $\beta$-model \citep{Cavaliere76}. Once the gas density profile has been set, gas temperatures follow from the requirement of hydrostatic equilibrium. The gas fraction is set at 15 per cent, which is a typical baryon content for the mass range of these objects \citep{Lagana2013}. Numerical realizations of these profiles are created with the techniques employed in \cite{Machado2013, Machado2015} and \cite{Ruggiero2019}. For further details, the reader is referred to those works and references therein. Each individual cluster is represented by $10^{6}$ gas particles and $10^{6}$ dark matter particles. 

% gadget
Once the initial conditions have been created, the collision is set up in the following manner. At $t=0$, the two clusters are placed 3\,Mpc apart along the $x$ axis and -- if needed -- also with an impact parameter $b$ along the $y$ axis. The initial relative velocity between the two clusters is $v_0$, parallel to the $x$ axis. It should be noted that the impact parameter refers to the $t=0$ configuration; the minimum separation, at the instant of pericentric passage, will be much smaller for typical velocities. Simulation was carried out with the hydrodynamical $N$-body code \textsc{Gadget-2} \citep{Springel2005} and the evolution was followed for at least 3\,Gyr. Given the spatial extent involved, cosmological expansion is ignored.

The purpose of the simulations is to offer a model that explains the current morphology of each cluster merger, allowing us to interpret its dynamical stage. A given simulation model will be considered a suitable representation of the cluster merger if it successfully recovers some observational constraints. Specifically, we are interested in reproducing mainly two observables: the temperature of the shock front and the separation between the two subclusters. The difficulty is in satisfying both criteria simultaneously. As the simulation evolves in time, an instant may be reached when both the temperature and the separation are approximately comparable to the observational data. This is what we will refer to as the `best instant' of each simulation: the moment in time when the simulations most closely resembles the observational results. From the simulation output, a sequence of simulated density and temperature maps is produced, at small time intervals. The best instant is determined by visual examination of these snapshots in comparison to the observed maps, and also by quantitative measurements of temperatures and distances. Once the best instant has been determined, one may indicate the absolute time $t$ since the beginning of the simulation, but this is an arbitrary time. It is physically more meaningful to express the best instant with reference to the instant of pericentric passage. From the simulation output, it is straighforward to identify the moment of pericentric passage. Therefore, in what follows, we will be able to determine the instant when the simulation best matches the observations, and we will express this moment in terms of the interval of time before the pericentric passage.

%%%%%%%%%%%%%%%%%%%%%%%%%%%%%%%%%%%%%%%%%%%%%%%%%%%
\section{Results and Discussion}
\label{res}

In this section we present the results for temperature profiles. To do that, we extracted spectra from rectangular regions (see Figs.~\ref{fig:caixaG292} and \ref{fig:caixaG036}) and fit them to single thermal models as described in Section~\ref{sec:XMM}. Also, to better characterise the thermodynamics and local variations we present 2D maps that  are shown in Figs.~\ref{fig:mapaG292} and \ref{fig:mapaG036}. 
% mach
 During the merger process of the subclusters, supersonic motion is produced in the ICM, consequently generating the shock front \citep[e.g.][]{Markevitch2002}. With the help of temperature profiles, it is possible to estimate the Mach number $\mathcal{M}$ from the amplitude of the temperature discontinuity \citep{Landau1959} $T_2/T_1$, using the equation:
\begin{equation}
\frac{T_2}{T_1} = \frac{5\mathcal{M}^4 + 14\mathcal{M}^2 - 3}{16\mathcal{M}^2},
\end{equation}
where an adiabatic index $\gamma = 5/3$ was assumed and,  the subscripts 1 and 2 denotes the regions pre-shock and post-shock, respectively. Thus, the uncertainties in the temperature ratio implies in the uncertainties of Mach number.

To investigate the morphologies of the subsystems, we fit a 2D spherical $\beta$-model \citep{Cavaliere76} to their surface brightness and, to take into account the ellipticity of the plasma emission we use the following standard coordinates transformation \citep[as done in][]{AndradeSantos12}: 

\begin{equation}
\begin{cases}
x^{\prime} =(x - x_{0}) \cos \theta  -  (y- y_{0}) \sin\theta \\
y^{\prime}  =(x - x_{0}) \sin \theta  -  (y- y_{0}) \cos \theta \\
r^{2}  = x^{\prime 2}  +  \frac{y^{\prime 2}}{(1-\epsilon)^{2}} &
\end{cases}
\end{equation}
where $(x_0,y_0)$ is the X-ray emission center coordinates for each X-ray peak, $\theta$ is the position angle, and $\epsilon$ is the ellipticity. The $\beta$-model may now be defined as follows:

\begin{equation}
\label{eq_beta}
S(r) = S_{0} \bigg(1+\frac{r^{2}}{r_{\rm c}^{2}}\bigg)^{-3\beta+1/2}  + b,
\end{equation}
where $S_{0}$ is the central brightness, $r_{\rm c}$ the core radius, $\beta$ the shape parameter, and $b$ a residual background emission, assumed
to be constant across the FOV. We assumed two $\beta$-models, each of them centered in the X-ray peak of the subsystems. Although this model should be an idealised representation of the surface brightness of a merging cluster, we are interested in the residual image that was obtained by subtracting the best 2D $\beta$-model from the original X-ray image.  

In order to model the clusters analysed in this work through numerical simulations, initial conditions were created with the parameters given in Tables~\ref{tab1} and \ref{tab2}. Observationally, the masses of the clusters were estimated within a certain radius $R$. We aimed to create clusters in such a way that their mass within the given radius, $M(<R)$, matched the observational estimates, at least to a rough approximation. Several simulation setups were explored until a sufficiently acceptable result was obtained. Here we report exclusively on the so-called best model of each cluster (G292 and G036).

%%%%%%%%%%%%%%%%%%%%%%%%%%%%%%%%%%%%%%%%%%%%%%%%%%%
\subsection{X-ray Analysis of G292}

\begin{figure}
\begin{center}
\includegraphics[width=0.8\columnwidth]{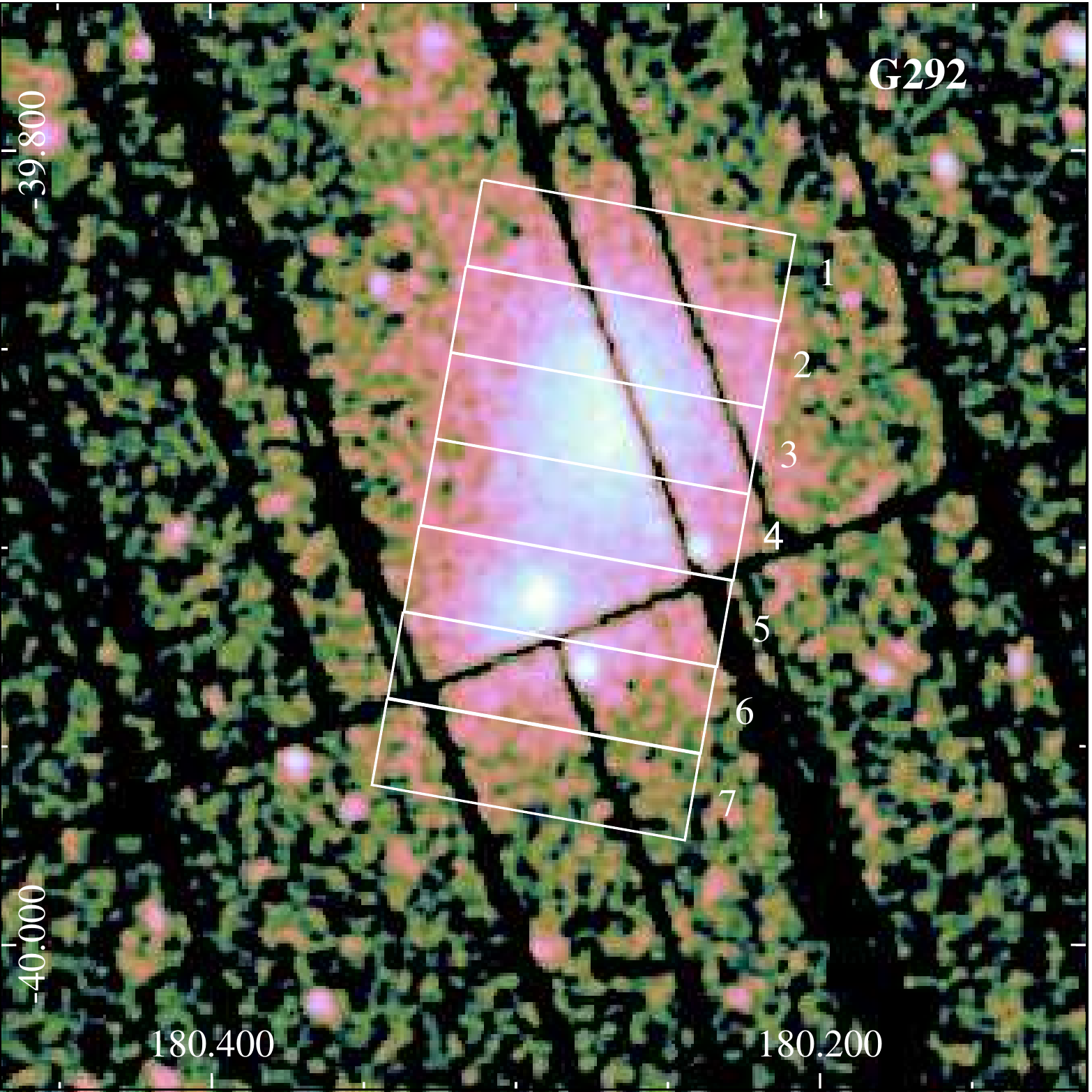}\\~\\
\includegraphics[width=\columnwidth,angle=180]{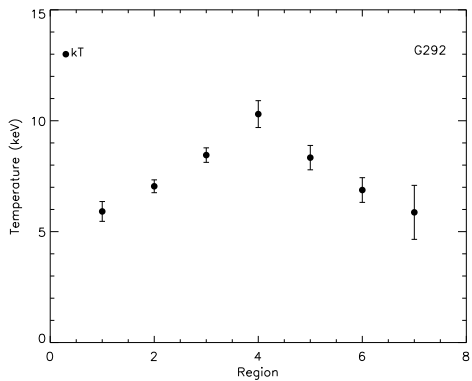}\\
\end{center}
\caption{Upper panel shows X-ray image for PLCK G292 with overlaid boxes used to determine temperature profile, shown in the lower panel.}
\label{fig:caixaG292}
\end{figure}

In Fig.~\ref{fig:caixaG292}  (upper panel) we show the regions used to obtain the temperature profile. The temperature profile indicates that in region 4, the temperature can be as high as 10 keV, suggesting a shock. If we assume region 2 as the pre-shock (with $T_{1}=7.04 \pm 0.29$ keV) and region 4 as the post-shock region (with $T_{2}=10.30 \pm 0.61$ keV), we obtain a Mach number of $\mathcal{M}=1.47 \pm 0.08$ for G292.

%--------------------------------------------------
%--------------------------------------------------

\begin{figure*}
\includegraphics[scale=0.3]{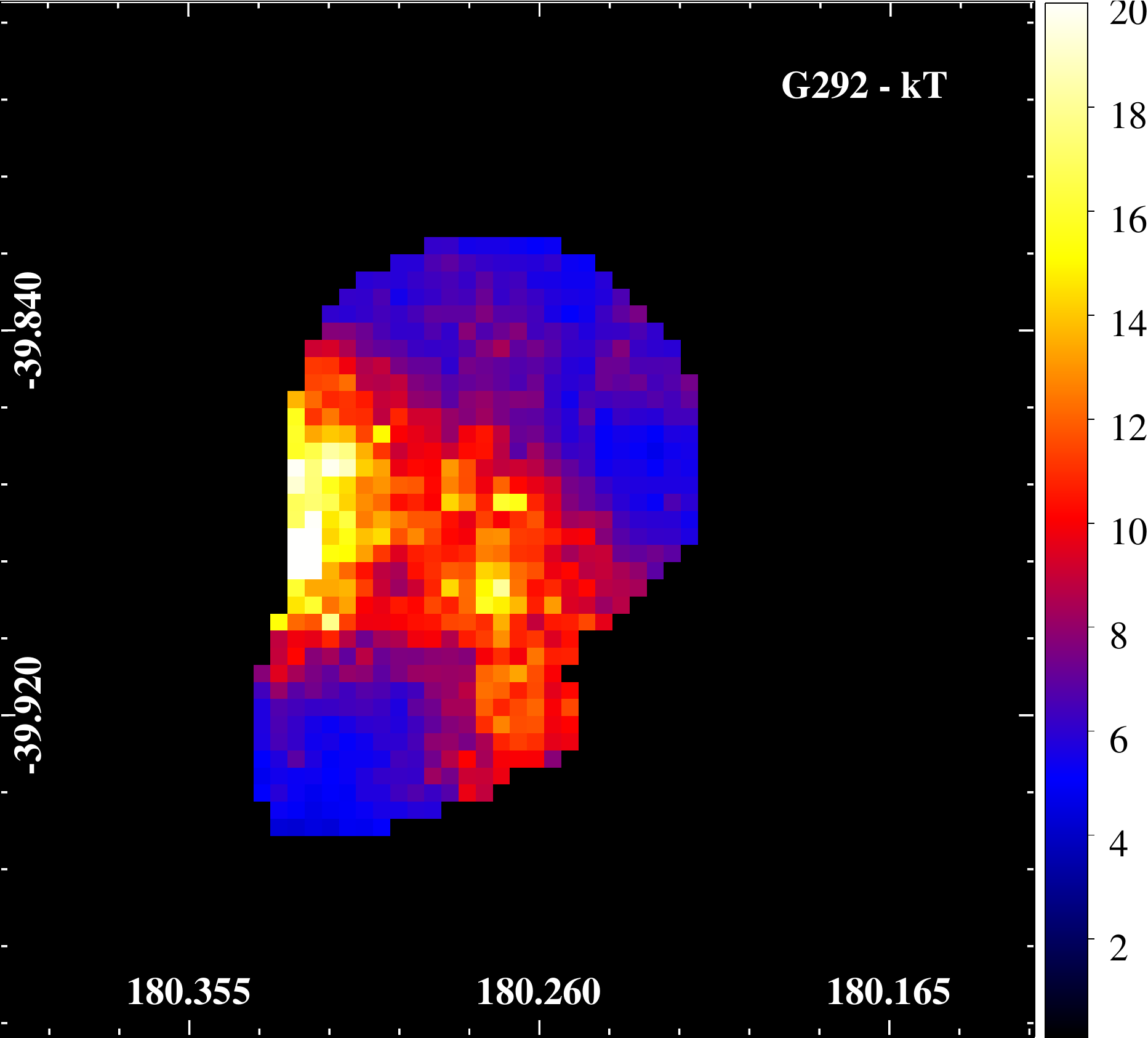}
\includegraphics[scale=0.3]{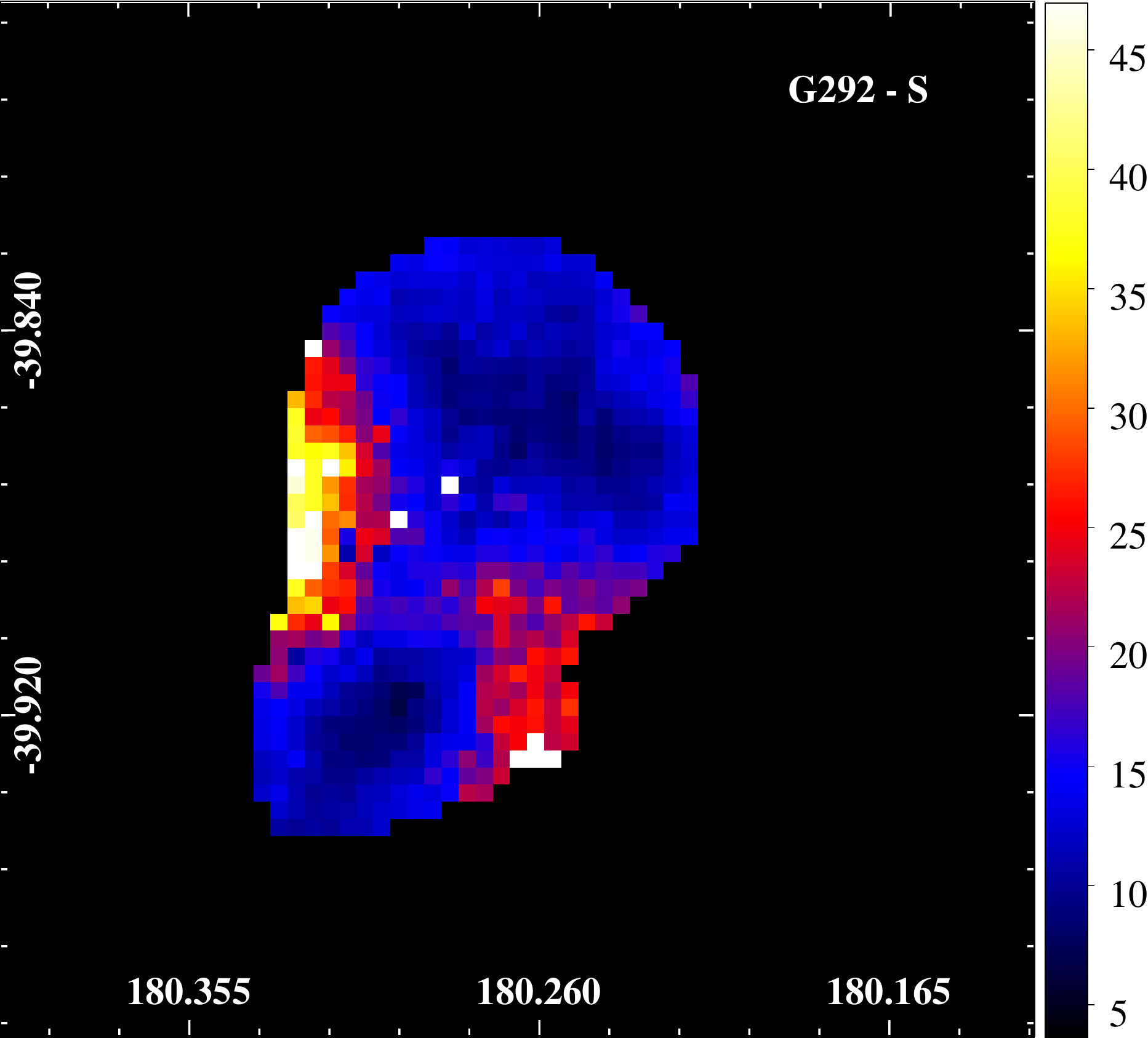}
\includegraphics[scale=0.3]{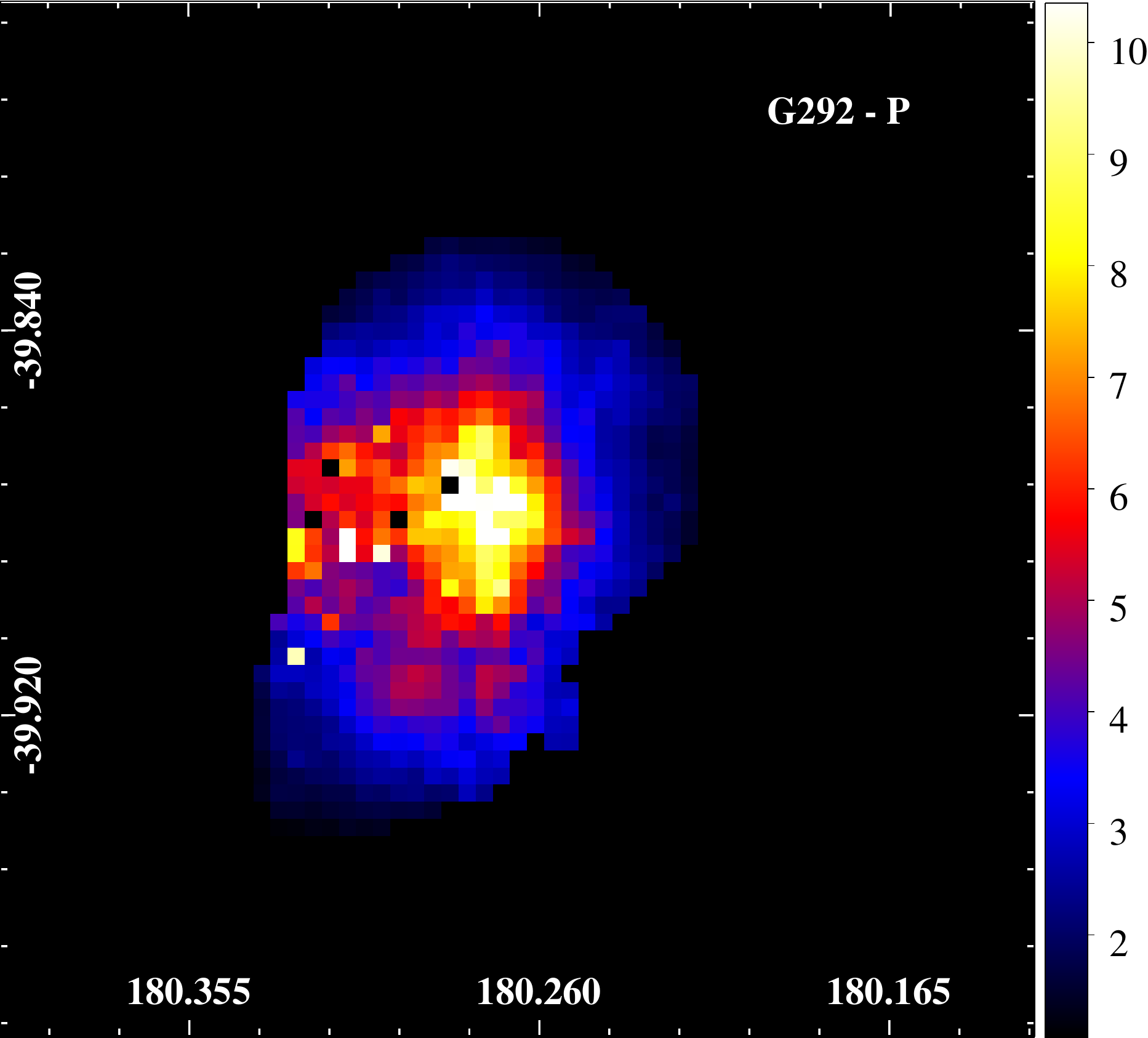}
\caption{2D maps for G292. Temperature (left), entropy (middle), and pressure (right) maps. Colour bars give values of the quantities. keV for temperature, and arbitrary units for
entropy and pressure. Coordinates are for J2000.}
\label{fig:mapaG292}
\end{figure*}

In Fig.~\ref{fig:mapaG292} we show the 2D spectral maps. The temperature map reveals that the ICM temperature varies substantially within the scale of the cluster, indicating its complex nature.  This hot region appears to be inomogeneously extended along the east-west direction likely due to the interactions between G292N and G292S. This  defining features of the temperature map reveal one of hottest shocks reported in the literature.

Entropy is the key parameter that records gain of the thermal energy through the shocks and/or AGN feedback while remaining insensitive to the adiabatic compressions and expansions. Its 2D map exhibits a significant increase in two regions at the outer edges of the shock. These high-entropy zones are spatially coincident with the two hottest areas in the $kT$ map. The pressure map, however, revealed a high-pressure center elongated almost perpendicularly to the shock front.

%--------------------------------------------------
%--------------------------------------------------

\begin{figure*}
\includegraphics[width=\textwidth]{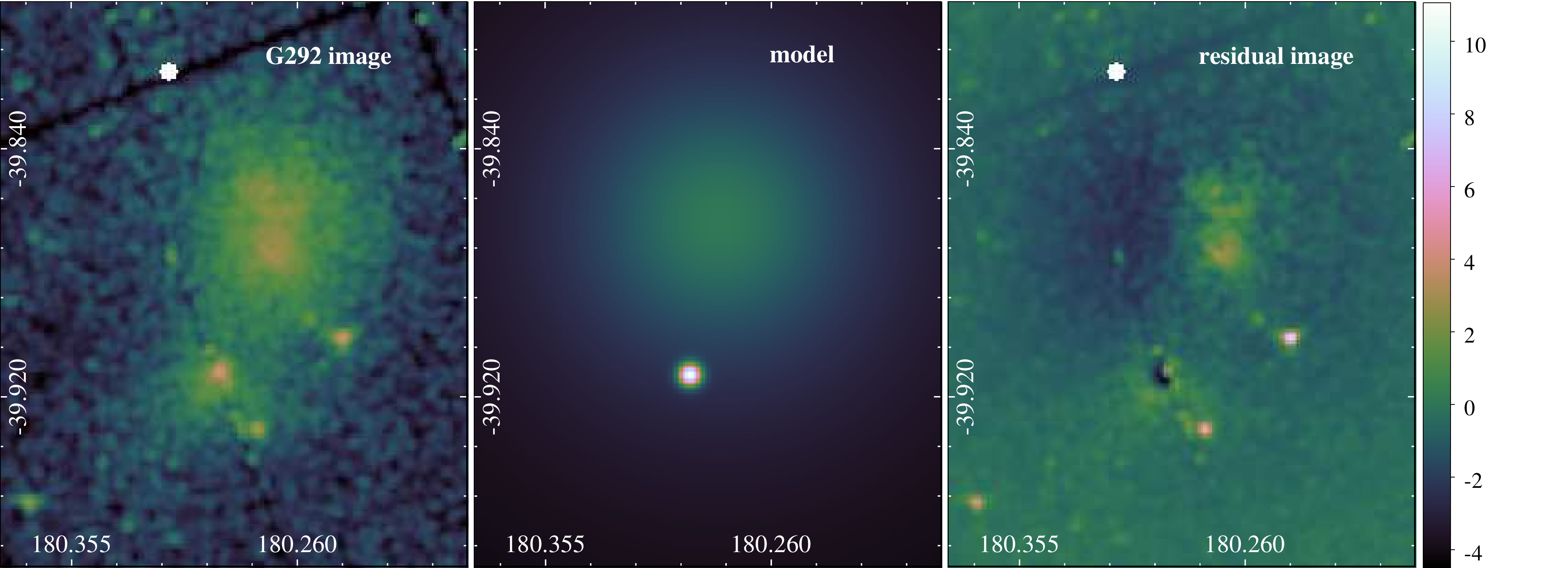}
\caption{X-ray surface brightness distribution (left), 2D $\beta$-model of the distribution (middle) and residuals (right) for G292, for which colorbar shows the intensity of the residue.}
\label{fig:residG292}
\end{figure*}

Besides the spectroscopic approach, we performed the 2D surface brightness fit.
The X-ray image, the best 2D $\beta$-model as well as the residual image are shown in Fig.~\ref{fig:residG292}. In the case of G292, the residual map does not evidence any substructure between the clusters. Some residuals are seen near the center of the subsystems, indicating deviations from spherical symmetry.

%--------------------------------------------------
%--------------------------------------------------

%%%%%%%%%%%%%%%%%%%%%%%%%%%%%%%%%%%%%%%%%%%%%%%%%%%
\subsection{Simulations of G292}

\begin{figure*}
\centering
\includegraphics[height=6cm]{figs/G292_RX.pdf}\hspace{5mm}
\includegraphics[height=6cm]{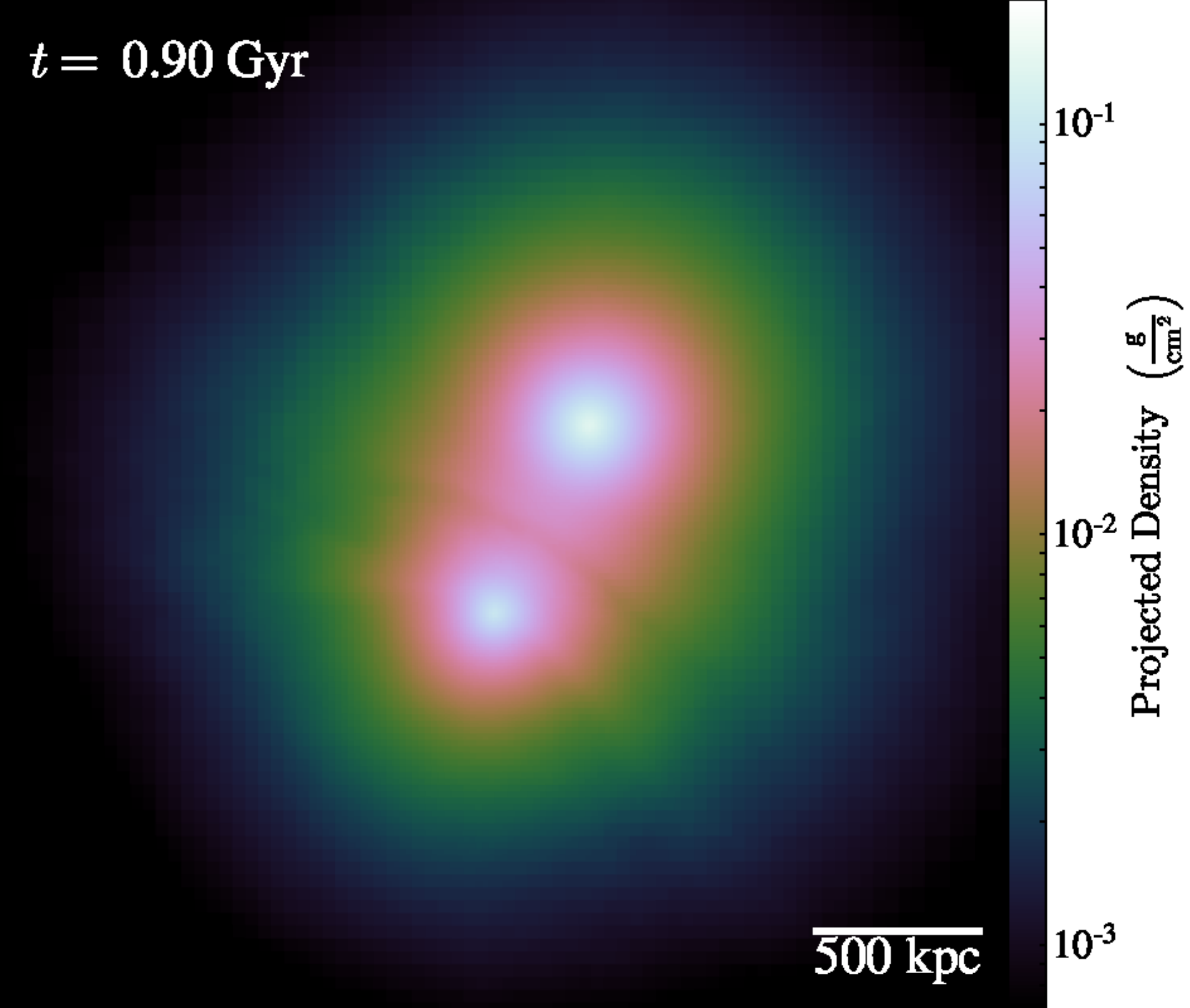}\\~\\
\includegraphics[height=6cm]{figs/G292_kTmap.pdf}
\includegraphics[height=6cm]{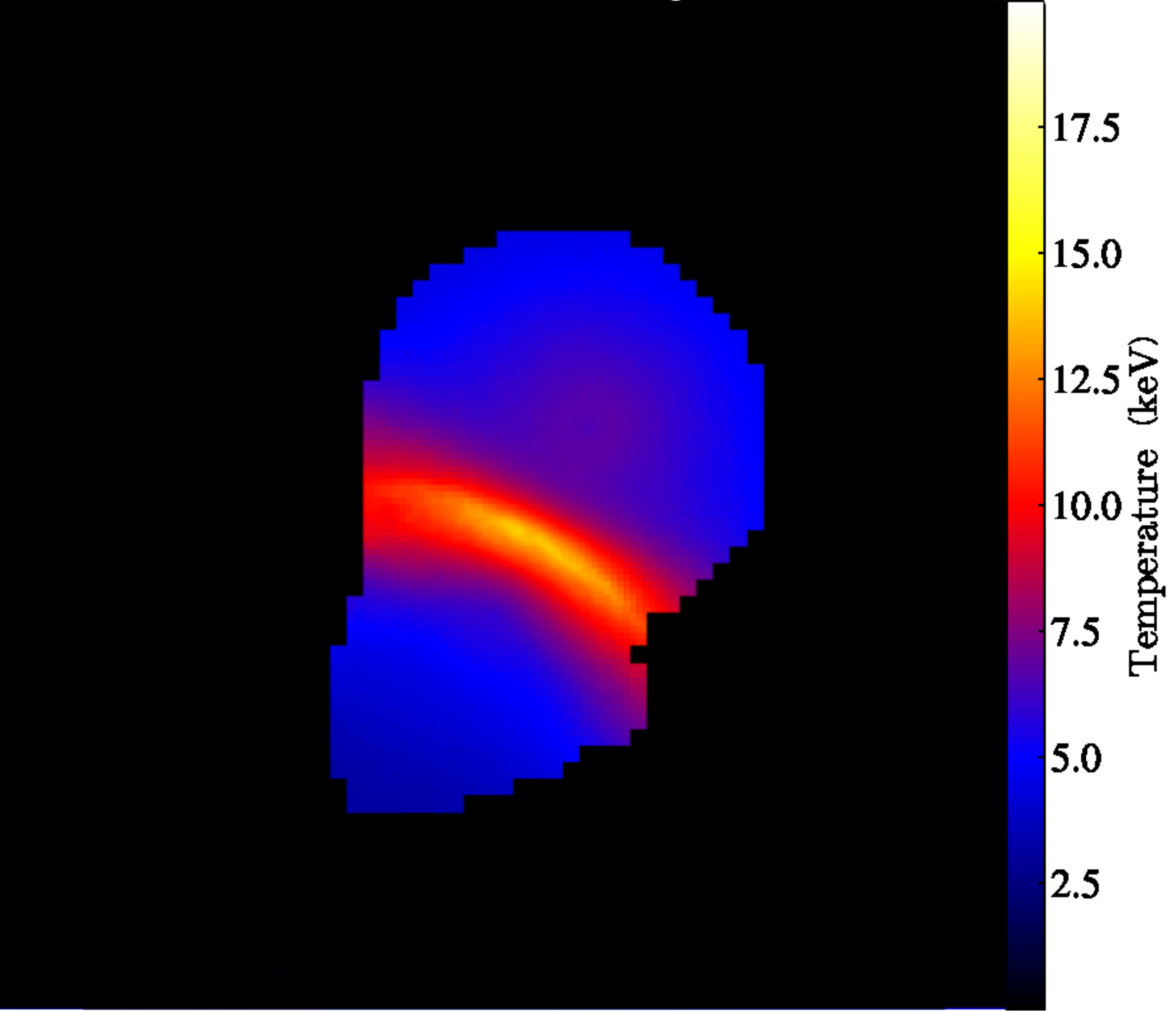}
\caption[]{Comparison between observations and simulations of G292. Left: observed X-ray emission and temperature map. Right: simulated projected density and simulated density-weighted projected temperature. The simulated temperature map was partially masked to match the corresponding region where observational results are available.}
\label{G292sim}
\end{figure*}

In an attempt to recover the morphology of G292, we performed several simulations of cluster mergers with the parameters given in Table~\ref{tab1}. A best simulation model was reached and it is shown in Fig.~\ref{G292sim}, where observations and simulation are compared side-by-side. In Fig.~\ref{G292sim}, the left panels are the observations and the right panels are the simulation at the best instant. The best instant was found to be $t=0.90$\,Gyr because this is the moment when the simulation snapshot matches the observational maps most closely. Prior to this instant, the temperature of the shock would still be insufficient and the separation would be too large. After this instant, the temperature of the shock would be excessive and the subclusters would be too close. For these reasons, $t=0.90$\,Gyr was judged to be the best instant of the simulation of G292.

% best instant
 For the case of G292, the best model was a frontal collision (impact parameter $b=0$) with initial relative velocity $v_0=-2000$\,km/s. We found that the best instant occurs at $t=0.90$\,Gyr. This is merely 150\,Myr prior to the central passage, which would take place at $t=1.05$\, Gyr. Thus the scenario is that of clusters that are coming in for their first approach. Given the uncertainties involved, it would seem unreasonable to ascribe much physical meaning to the specific value of this short interval. However, it is meaningful that the best moment is before central passage and not after. In a frontal collision, the morphology changes drastically after the cluster cores pass through each other. In fact, we found no plausible candidate model in which G292 could be the aftermath of a head-on encounter.

% temperatures
At the best instant, the collision axis needs to be inclined by $i=18^{\circ}$ with respect to the plane of the sky. This causes the projected separation between the clusters to be 643\,kpc (within 5 per cent of the observed separation of 677 kpc). The inclination angle is not tightly constrained, however. A range of roughly $5^{\circ}$ around the central value does not alter the projected morphology significantly and still produces tolerable results. Simultaneously, the temperature of the shock front is in the approximate range 11.4--13.6\,keV, whereas the pre-shock gas is in the range 6.6--7.2\,keV (Fig.~\ref{G292sim}), resulting in a fair quantitative agreement with the observed temperatures. In Fig.~\ref{G292sim}, the snapshot was rotated on the plane of the sky, simply in order to match the position angle of the observation; this rotation does not affect physical quantities. The North and South labels of Table~\ref{tab1} evidently refer to this configuration.

% inclination
The inclination angle $i$ (between the collision axis and the plane of the sky) does have meaningful consequences. It obviously affects the projected distance, but it also affects the projected temperature of the shock front. In a frontal collision, the central surface of the shock front is essentially perpendicular to the collision axis. If the collision axis lies on the plane of the sky ($i=0$), then the shock front is seen at its sharpest possible configuration (edge-on, so to speak) and the temperatures are quite high. Seen under a non-zero inclination $i$, more of the surrounding gas contributes along the line of sight, thereby attenuating the projected temperature. As a result, Mach numbers measured via the observed amplitude of the temperature jump are underestimated \citep[see e.g.][]{Machado2013}.

% mach
The Mach number was estimated from the simulation using the temperature jump across the shock front. With the simulated temperature map of Fig.~\ref{G292sim} (at $t=0.90$\,Gyr and already inclined by $i=18^{\circ}$), we measured a $T_{2}/T_{1}$ ratio which implies Mach numbers in the range $\mathcal{M} \sim 1.6 - 2.0$. This is in fair agreement with the observational estimate of $\mathcal{M}\sim1.5$. If the temperatures had been measured without inclination ($i=0^{\circ}$) at the same instant, the resulting Mach numbers would have been slightly higher, in the range $\mathcal{M} \sim 1.9 - 2.1$. Considering a short interval of 0.02\,Gyr around the best instant, the Mach numbers are constant to within a few per cent.

\begin{figure*}
\includegraphics[width=\textwidth]{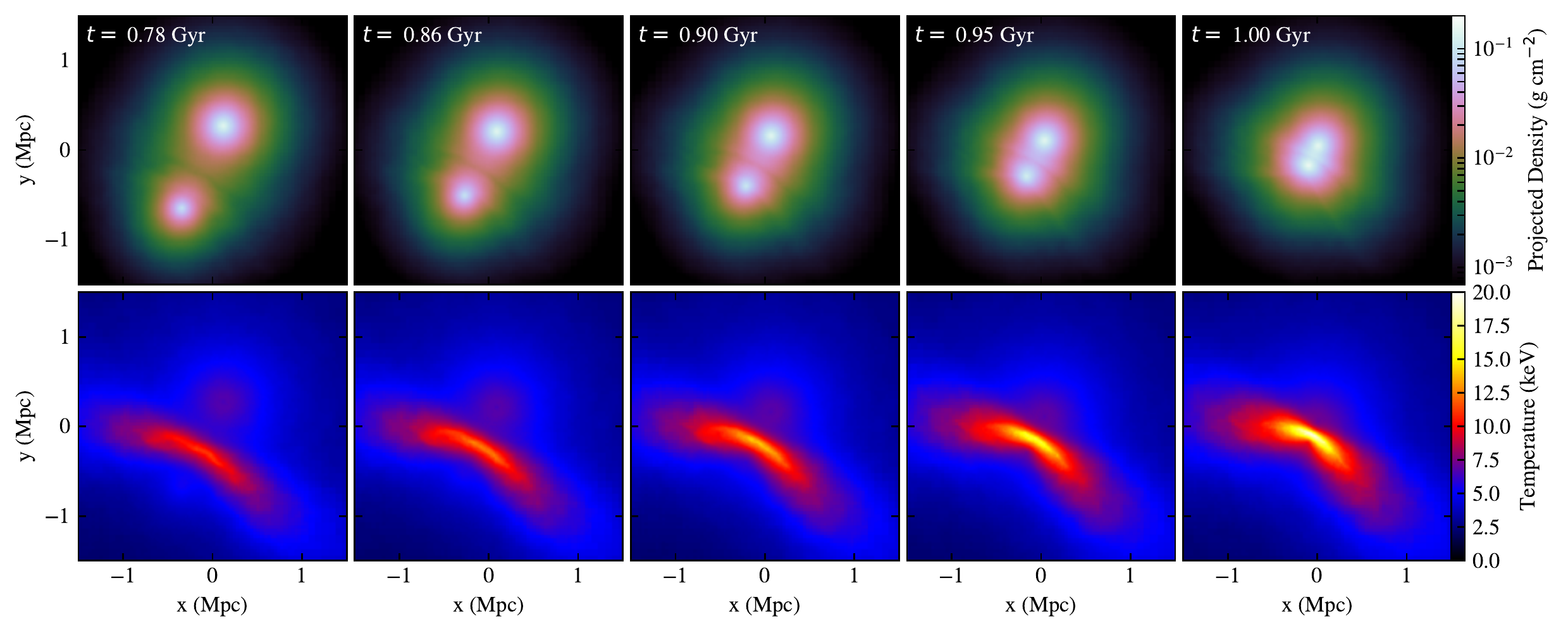}
\caption[]{Time evolution of simulated G292. Upper panels show projected density maps and lower panels show density-weighted projected temperature maps. The best instant corresponds to central panels ($t=0.90$\,Gyr).}
\label{G292time}
\end{figure*}

% temporal
A temporal evolution of the best model for G292 is shown in Fig.~\ref{G292time}, which displays densities and temperatures as a function of time. The panels are separated by steps of approximately 0.05\,Gyr. One notices that during this time span of nearly 0.25\,Gyr shown in Fig.~\ref{G292time}, the temperature of the shock front rises by about 5\,keV as the clusters approach each other.

% shortcomings
These simulations are naturally idealized and cannot be expected to account for all details of the observed objects. Even though this best model succeeds in reproducing quantitatively the masses, separation and temperature ranges of the gas, there are shortcomings. For example, the morphology of the simulated shock front is quite thin and, in particular, exhibits no hint of the excess in temperatures seen in the eastern region of the observed temperature map (see Appendix \ref{ApfontesG292}).

%--------------------------------------------------------------------
\begin{table}
\caption{Initial condition parameters for the G292 simulation. For the northern and southern clusters, this table gives: $M_{500}$, $r_{500}$, central temperature, the radius $R$ within which mass was measured, and the mass within that radius.}
\label{tab1}
\begin{center}
\begin{tabular}{l c c}
\hline
 & G292N & G292S \\
\hline
$M_{500}$ (${\rm M}_{\odot}$) & $8.0 \times 10^{14}$ & $3.6 \times 10^{14}$ \\
$r_{500}$ (kpc)               & 1913                 & 1467                 \\
$T$ (keV)                     & 6.5                  & 5.5                  \\
$R$ (kpc)                     & 670                  & 360                  \\
$M(<R)$ (${\rm M}_{\odot}$)   & $4.1 \times 10^{14}$ & $1.5 \times 10^{14}$ \\
\hline
\end{tabular}
\end{center}
\end{table}
%--------------------------------------------------------------------

%--------------------------------------------------------------------
%--------------------------------------------------------------------

%--------------------------------------------------------------------
%--------------------------------------------------------------------

%%%%%%%%%%%%%%%%%%%%%%%%%%%%%%%%%%%%%%%%%%%%%%%%%%%
\subsection{X-ray Analysis of G036}

%--------------------------------------------------
\begin{figure}
\begin{center}
\includegraphics[width=0.8\columnwidth]{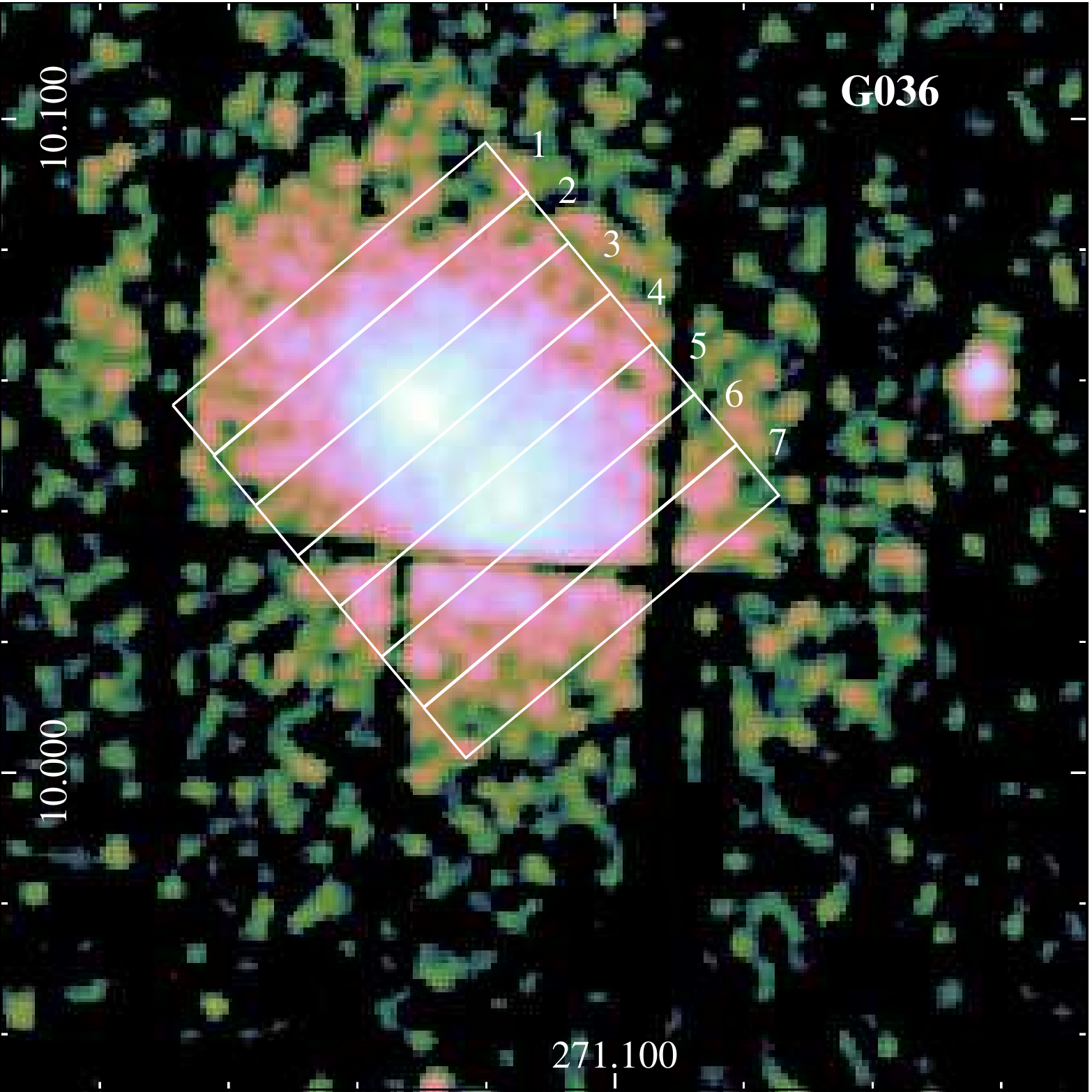}
\includegraphics[width=\columnwidth,angle=180]{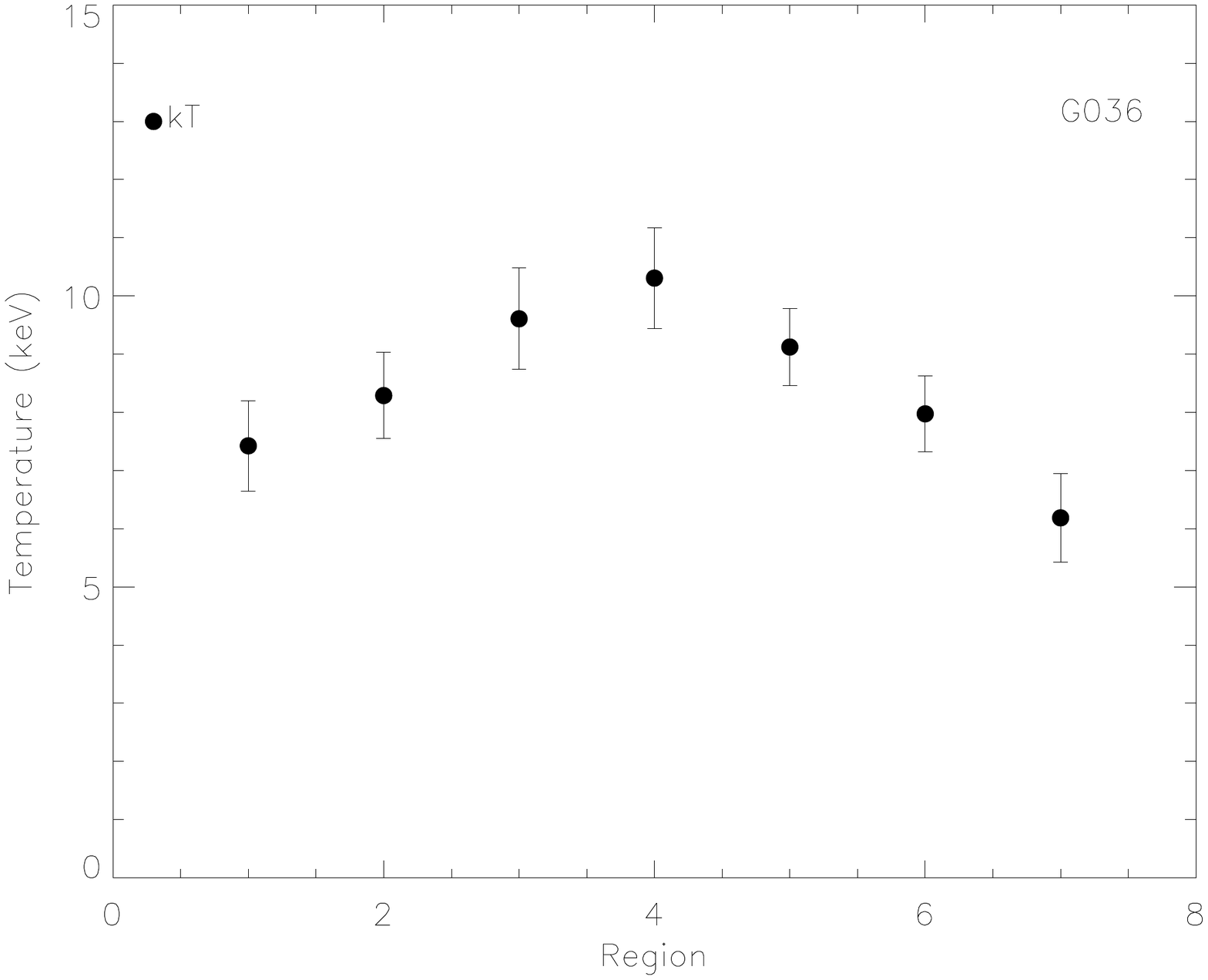}
\end{center}
\caption{Upper panel shows X-ray image for PLCK G036 with overlaid boxes used to determine temperature profile, shown in the lower panel.}
\label{fig:caixaG036}
\end{figure}
%--------------------------------------------------

From the temperature profile shown in the bottom panel of Fig.~\ref{fig:caixaG036}, we see that the temperature rises from around 8 keV in the regions that encompasses the centre of the subclusters (regions 3 and 5) to more than 10 keV in the region between the systems (consistent with uncertainties), indicating a shock due to the merger. For G036, from Fig.~\ref{fig:caixaG036} we assumed region 2 as the pre-shock, with a temperature of $T_{1}=8.29 \pm 0.74$ keV, and region 4 as the post-shock  with a temperature of $T_{2}=10.31 \pm 0.61$ keV. We obtained a Mach number $\mathcal{M} = 1.3 \pm 0.1$ for this merging cluster, which is in line with the weak shock ($\mathcal{M} \sim $ 1.0--1.6)  obtained by \citet{Zhang15} using XMM-\textit{Newton} data for this cluster.

To obtain the residual image, we subtracted  a 2D $\beta$-model (equation~\ref{eq_beta}) from the main cluster emission. Since G036S and G036N are very close in projection (178 kpc), the model appears to have one component, but it is the sum of two $\beta$-models centered on the emission peaks. The residual map reveals
perturbations between the subsystems, corroborating a scenario of a shock front due to the merger process. Also, we still see some residuals associated with the centers of G036S and G036N.

%--------------------------------------------------
\begin{figure*}
\includegraphics[scale=0.3]{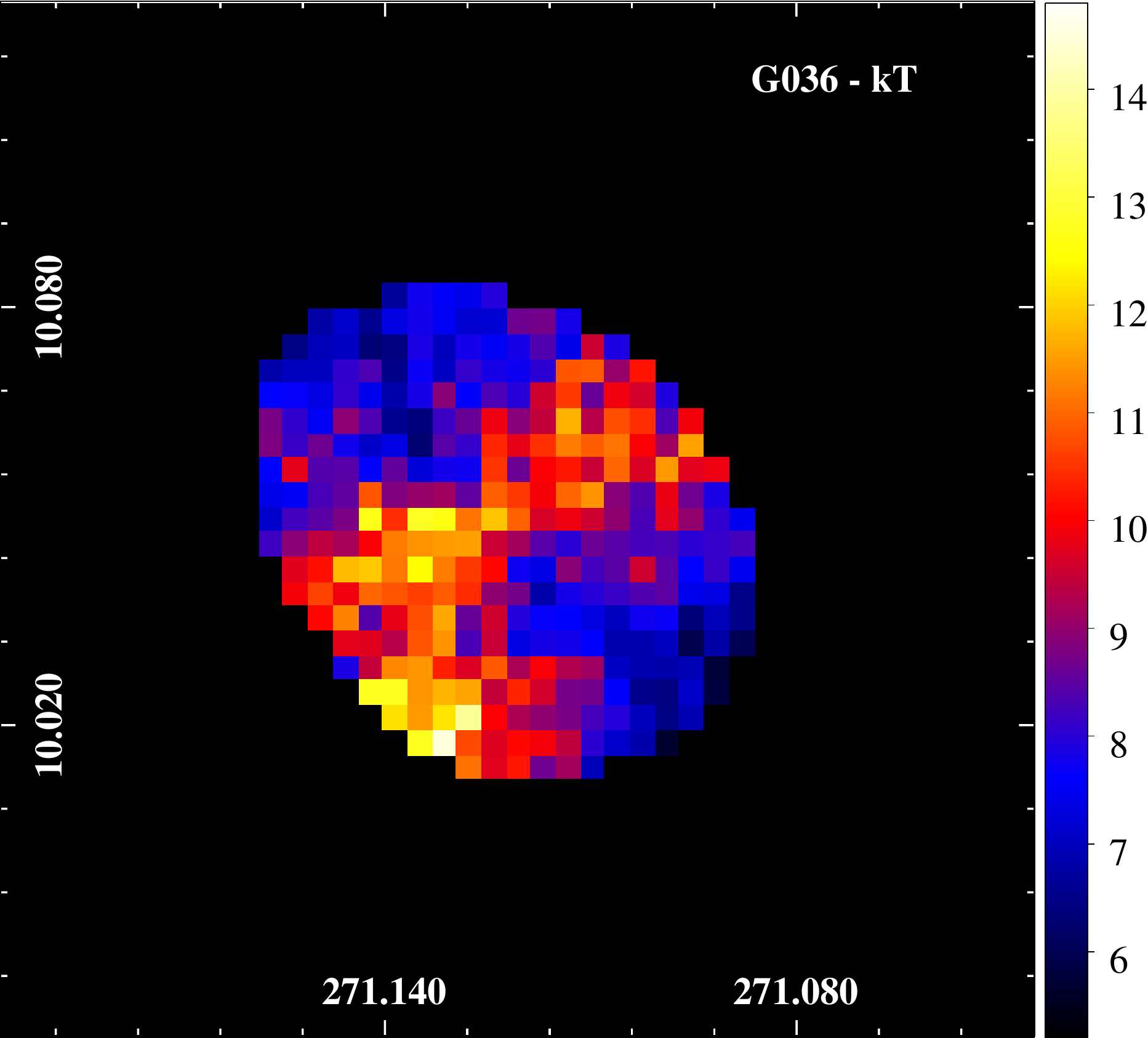}
\includegraphics[scale=0.3]{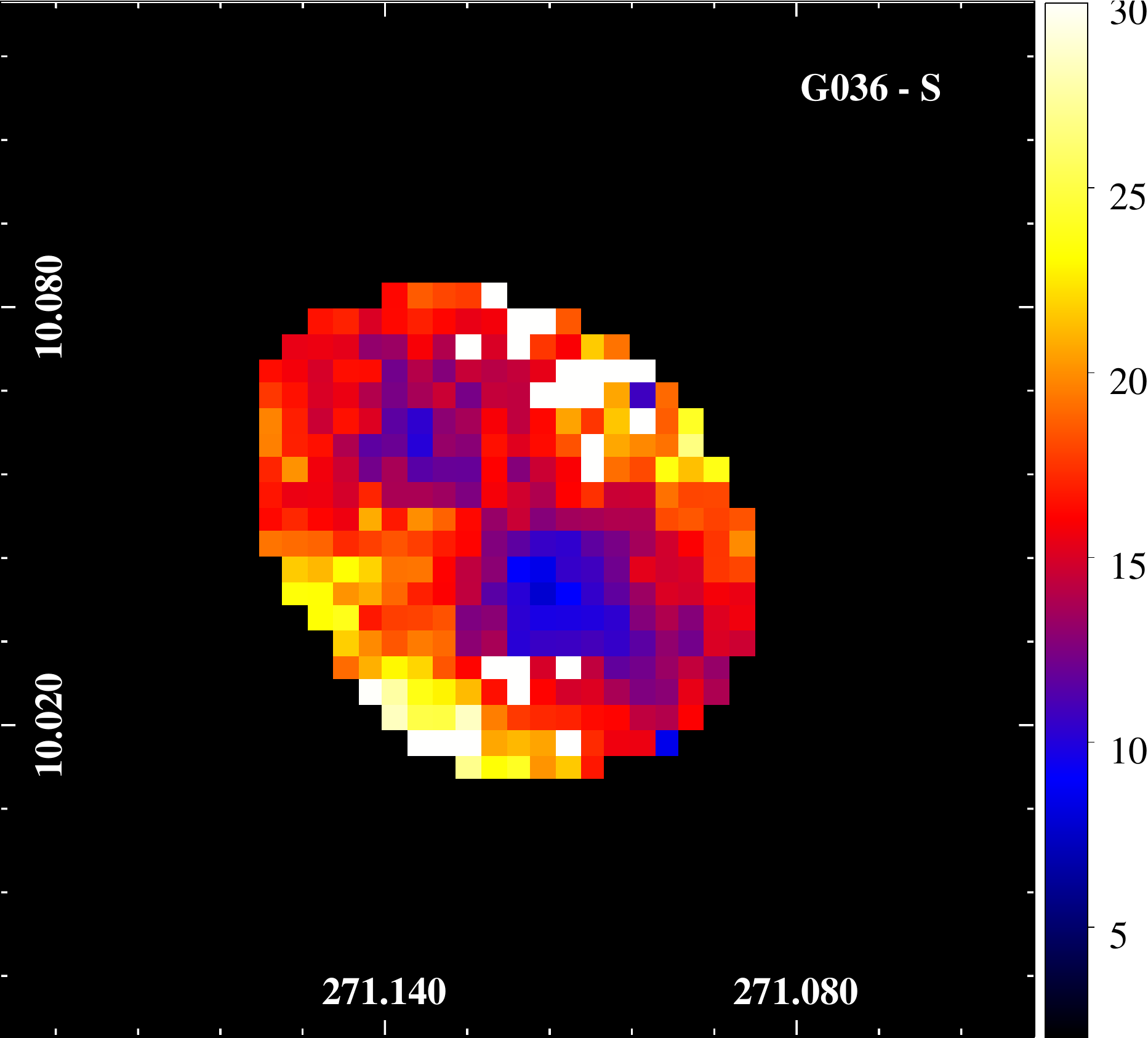}
\includegraphics[scale=0.3]{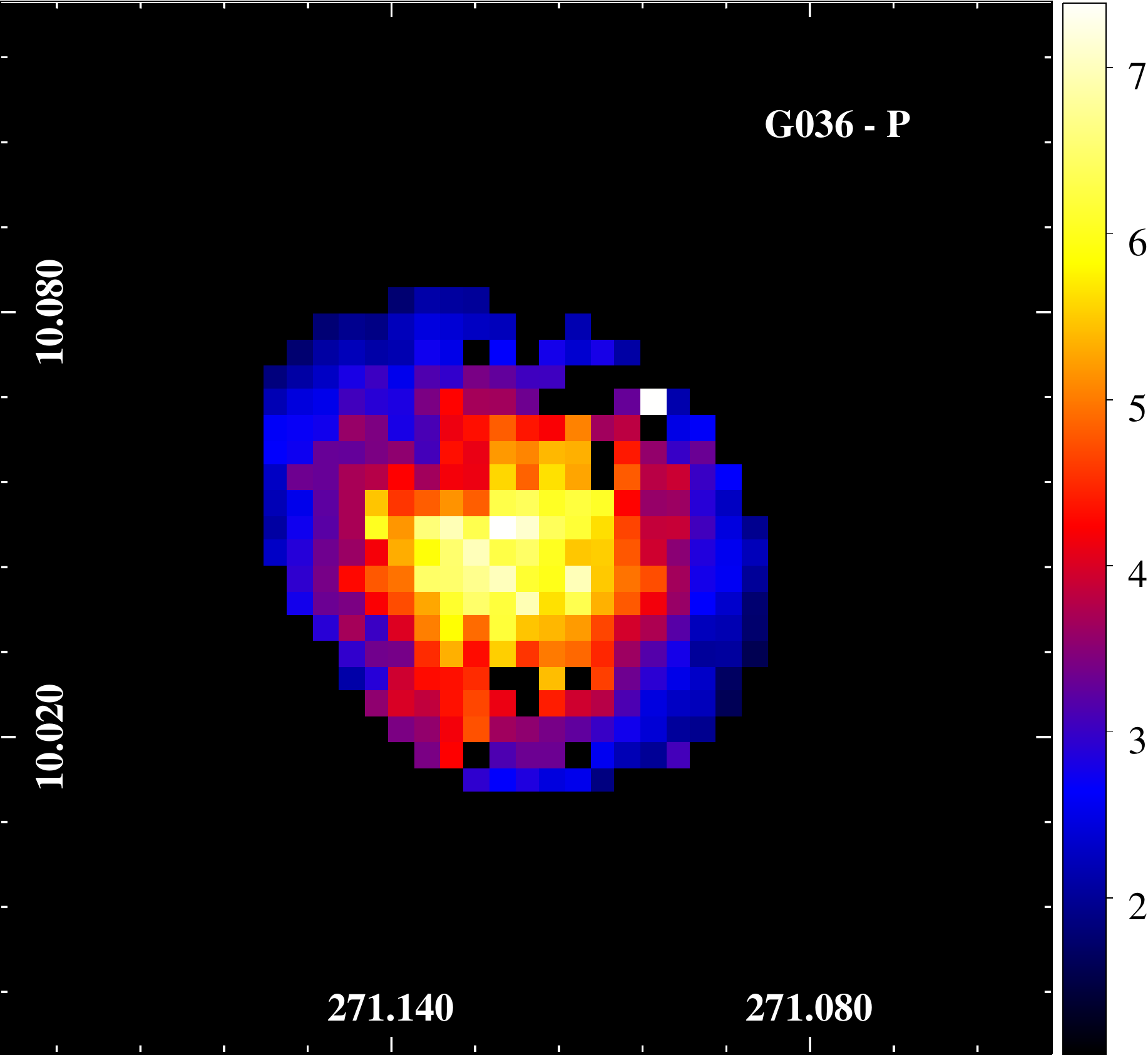}
\caption{2D maps for G036. Temperature (left), entropy (middle), and pressure (right) maps. Colour bars give values of the quantities. keV for temperature, and arbitrary units for
entropy and pressure. Coordinates are for J2000.}
\label{fig:mapaG036}
\end{figure*}   
%--------------------------------------------------

%--------------------------------------------------
\begin{figure*}
\includegraphics[width=\textwidth]{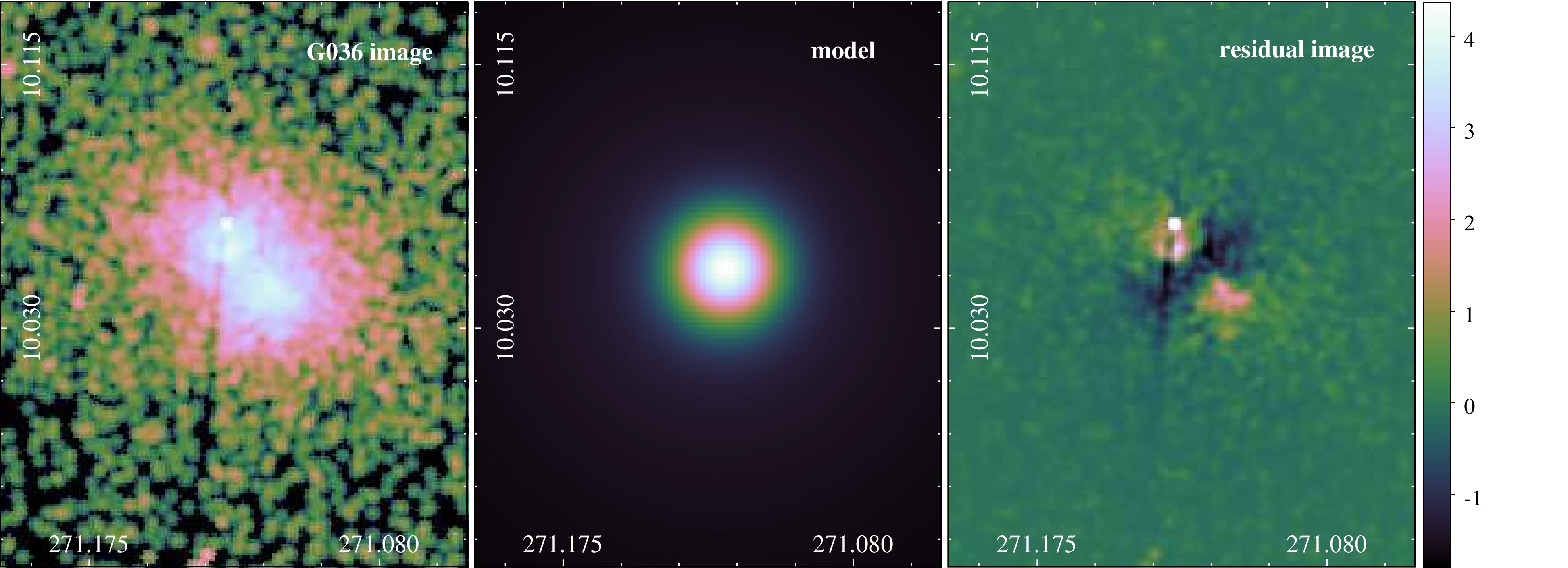}
\caption{X-ray surface brightness distribution (left), 2D $\beta$-model of the distribution (middle) and residuals (right) for G036, for which colorbar shows the intensity of the residue.}
\label{fig:residG036}
\end{figure*}  
%--------------------------------------------------
 
To better analyse this merging cluster, we show in Fig.~\ref{fig:mapaG036} the spatial distribution of $kT$, $P$, and $S$. Merger shocks can be observed as temperature, pressure and entropy discontinuities in the hot X-ray gas. The temperature map confirms the results found in the temperature profile of a high-temperature region between the two centroids (with $kT \sim$ 10--11 keV). The entropy map shows two regions of lower entropy, coincident with the position of the two subsystems. The region between G036S and G036N shows higher entropy values (a factor of 2 compared to the X-ray peaks regions), corroborating a scenario of a shocked region due to an ongoing merger \citep[as in][]{Zhang15}. The pressure map reveals an inner region, almost spherically symmetric, of high pressure centered between the subsystems.

%%%%%%%%%%%%%%%%%%%%%%%%%%%%%%%%%%%%%%%%%%%%%%%%%%%
\subsection{Simulations of G036}

% parameters
Similarly to the previous case, here we present the results from the simulations of G036. Table~\ref{tab2} gives the initial parameters of the best model.

%--------------------------------------------------------------------
\begin{table}
\caption{Initial condition parameters for the G036 simulation. For the northern and southern clusters, this table gives: $M_{500}$, $r_{500}$, central temperature, the radius $R$ within which mass was measured, and the mass within that radius.}
\label{tab2}
\begin{center}
\begin{tabular}{l c c}
\hline
 & G036N & G036S \\
\hline
$M_{500}$ (${\rm M}_{\odot}$) & $3.1 \times 10^{14}$ & $3.6 \times 10^{14}$ \\
$r_{500}$ (kpc)               & 1028                 & 1036                 \\
$T$ (keV)                     & 8.5                  & 9.0                  \\
$R$ (kpc)                     & 105                  & 105                  \\
$M(<R)$ (${\rm M}_{\odot}$)   & $0.7 \times 10^{14}$ & $0.9 \times 10^{14}$ \\
\hline
\end{tabular}
\end{center}
\end{table}
%--------------------------------------------------------------------

\begin{figure*}
\centering
\includegraphics[height=6cm]{figs/G036_RX.pdf}\hspace{5mm}
\includegraphics[height=6cm]{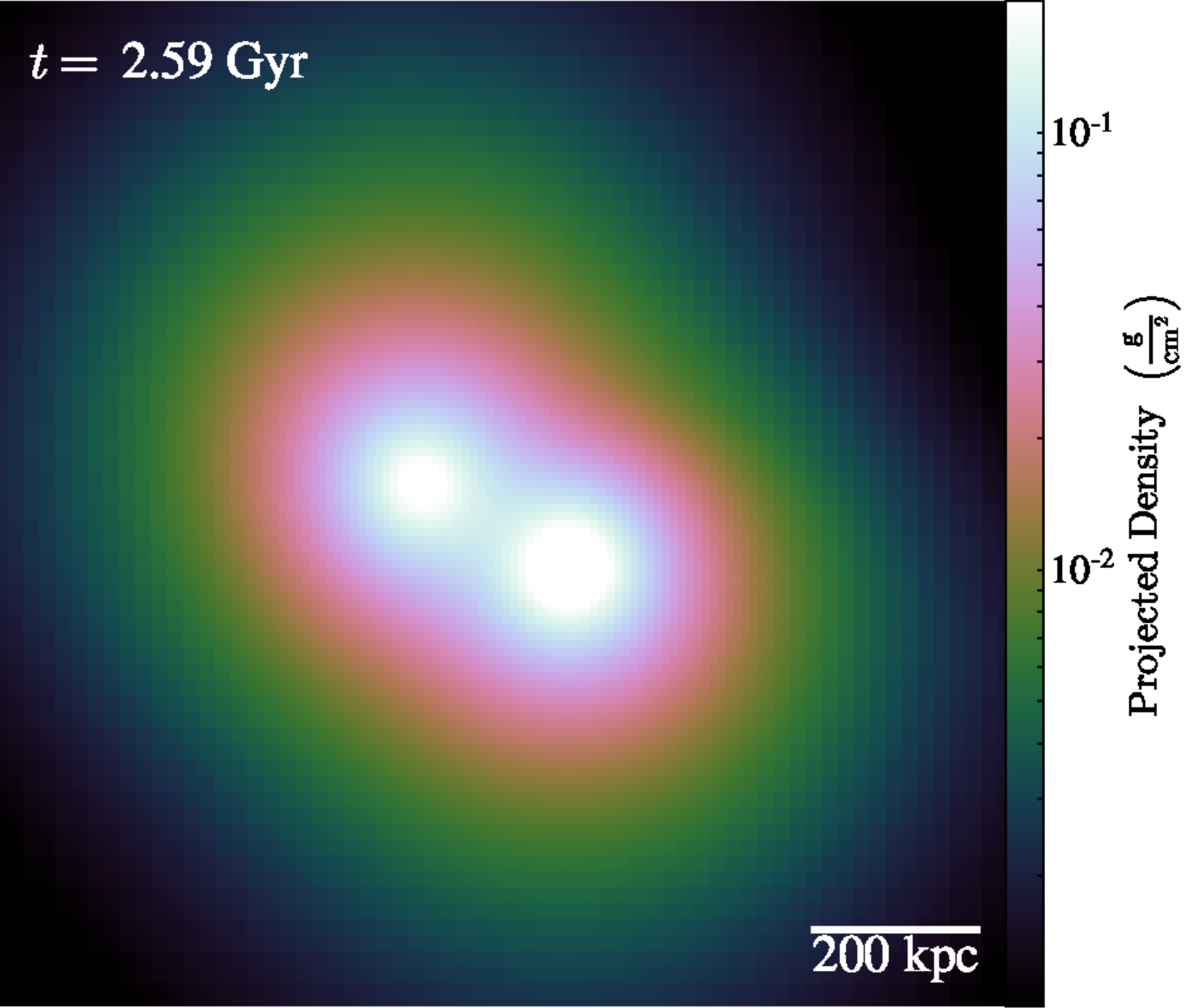}\\~\\
\includegraphics[height=6cm]{figs/G036_kTmap.pdf}
\includegraphics[height=6cm]{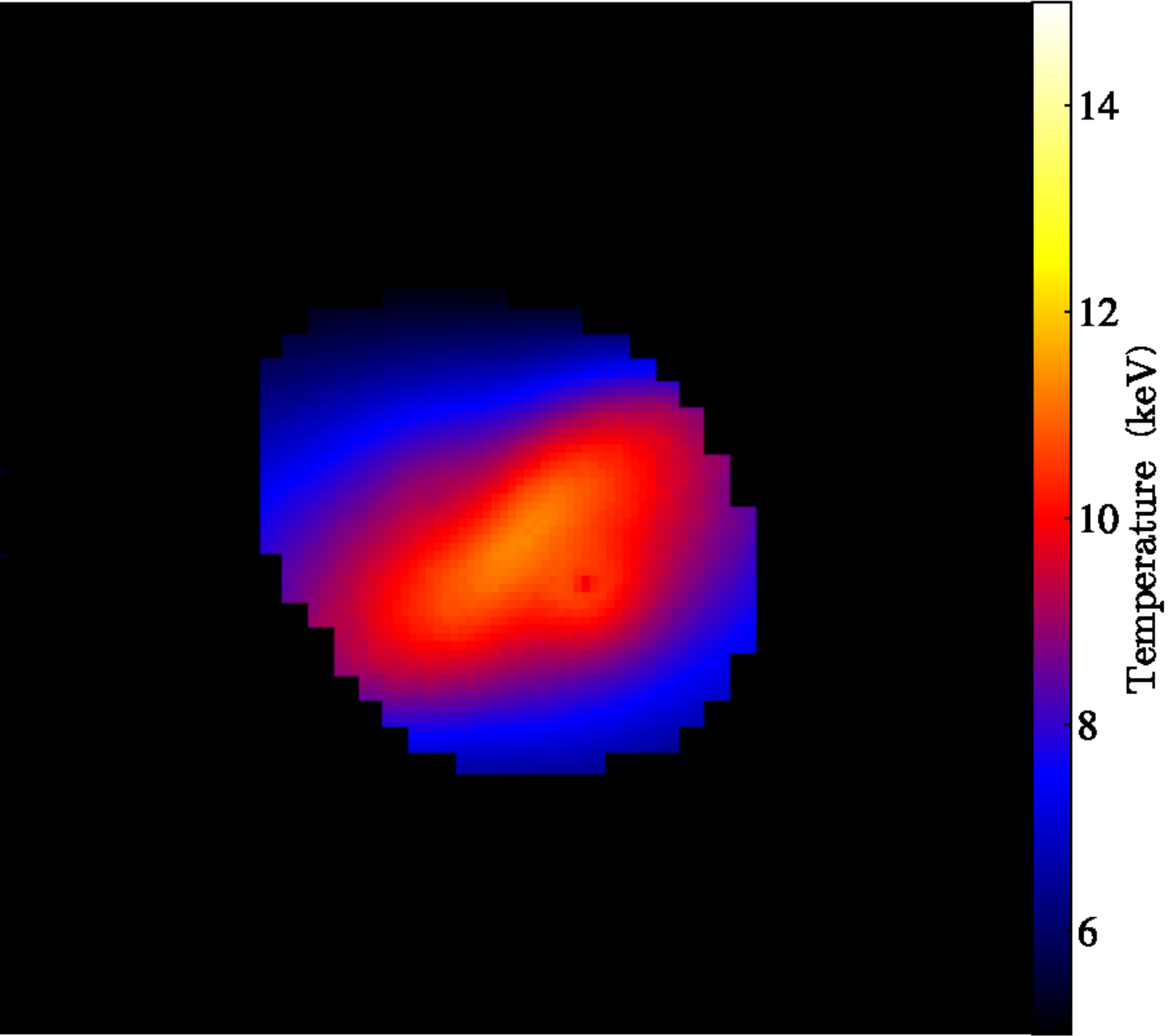}
\caption[]{Comparison between observations and simulations of G036. Left: observed X-ray emission and temperature map. Right: simulated projected density and simulated density-weighted projected temperature. The simulated temperature map was partially masked to match the corresponding region where observational results are available.}
\label{G036sim}
\end{figure*}

% best instant
For the case of G036, the best model is not a frontal collision, but starts with an impact parameter of $b=1600$\,kpc, with initial relative velocity $v_0=-700$\,km/s. We found that the best instant occurs at $t=2.59$\,Gyr, merely 50 Myr prior to the pericentric passage, which would take place at $t=2.64$\, Gyr. Again in the case of G036, the best instant of the simulation is the moment when the separation between the subcluster matches the observed separation, and the temperature of the shock is approximately in the same range as the observed temperatures. The best instant is shown in Fig.~\ref{G036sim}.

% inclination
In order to match the observed serparation of 178\,kpc, the plane of the orbit must be inclined with respect to the plane of the sky by $i=50^{\circ}$ in this case. Such inclination also renders the temperature range acceptable at the same time: the shock temperature would be roughly 9.5--10.8\,keV, whereas the pre-shock gas would be in the range 6.6--7.4\,keV (Fig.~\ref{G036sim}).

In this model, if the inclination angle was smaller, the projected temperature of the shock front would be excessive and the projected separation between the subclusters would be too large. If the orbital plane was even more inclined, however, then the projected temperature would be attenuated and the two subclusters would appear too close in projection. Therefore, the most suitable inclination was determined to be $i=50^{\circ}$ at this best instant, since it provided an acceptable compromise in approximately satisfying both criteria. \cite{Zhang15} had previously estimated a considerably large inclination angle of roughly $80^{\circ}$ for this system.

% mach
The simulated Mach number of G036 was similarly obtained using the temperature jump measured from the temperature map of Fig.~\ref{G036sim} (at $t=2.59$\,Gyr and inclined by $i=50^{\circ}$). The temperature ratio resulted in Mach numbers of approximately $\mathcal{M}\sim1.3 - 1.6$, which is consistent with the observational estimate of $\mathcal{M}\sim1.3$. With no inclination ($i=0^{\circ}$), the Mach numbers would have been in the range $\mathcal{M} \sim 1.5 - 2.0$ at the same instant.

\begin{figure*}
\includegraphics[width=\textwidth]{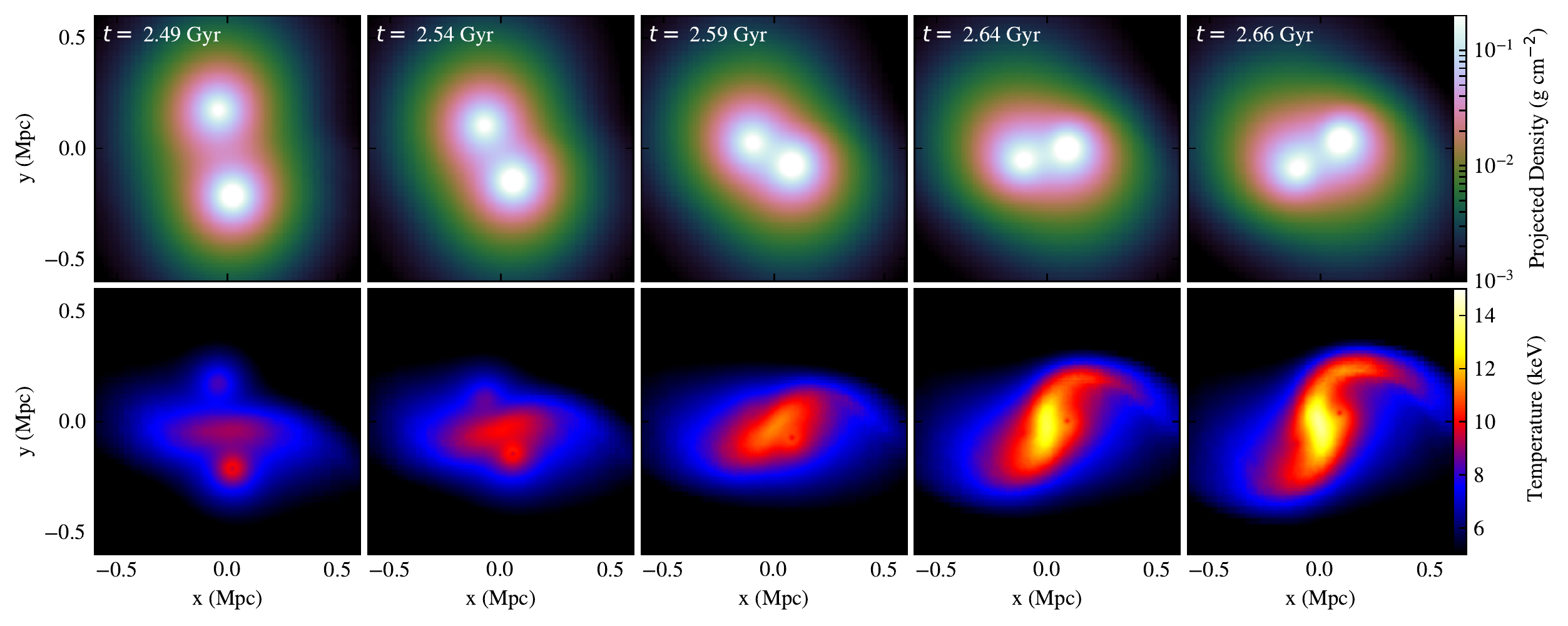}
\caption[]{Time evolution of simulated G036. Upper panels show projected density maps and lower panels show density-weighted projected temperature maps. The best instant corresponds to central panels ($t=2.59$\,Gyr).}
\label{G036time}
\end{figure*}

% temporal
Figure~\ref{G036time} displays the temporal evolution of the simulated G036. In contrast to the time evolution of G292, this sequence of snapshots is less intuitive to follow visually, for two reasons. First, there is a considerable inclination ($i=50^{\circ}$) between the plane of the orbit and the plane of the sky. Second, and more importantly, the collision is not frontal, meaning that the clusters are not falling radially towards each other. Nevertheless, examining by eye the motion of centroids of the density maps, one may surmise that the pericentric passage must occur between the third and fourth panels, and that the cores do not interpenetrate. The changes in the temperature structure in the corresponding maps indicate a complicated evolution after pericentric passage (not shown). Indeed, at times later than the ones shown, the spiraling morphology of the hot gas would become even more complicated and it could hardly be referred to as a shock front thenceforth. The final frame of Fig.~\ref{G036time} already indicates that the subsequent evolution would depart significantly from the target morphology.

% shortcomings
Also in the case of G036, not all observed morphological features are captured by the simulation. Here again, the masses, separation and temperature ranges are sufficiently accomodated. However, the simulation fails to recover the detailed  morphology of the shock front, which displays a substantially curved shape for the southeastern edge in the observational data.

%--------------------------------------------------------------------
%--------------------------------------------------------------------

%--------------------------------------------------------------------
%--------------------------------------------------------------------

% What is most significant about these results is that the predicted X-ray surface brightness and  temperature at about XXGyr before closest approach agree remarkably well with the observed properties observed in X-ray.

%%%%%%%%%%%%%%%%%%%%%%%%%%%%%%%%%%%%%%%%%%%%%%%%%%%
\section{Conclusions}
\label{conc}
%X-rays
With our high resolution  XMM-\textit{Newton} observations,  we analysed two merging clusters (G036 and G292) through spectral 2D maps and X-ray surface 
brightness residual images. Although G036 had two subsystems very close in projection, they are seen as separated systems.
Temperature profiles were extracted in regions along the projected merger direction and indicated high temperature values between the subsystems of both G036 and G292. The temperatures reached $\sim 10-12$ keV (in both G036 and G292), indicating very hot shock fronts, some of the hottest reported in the literature.
Our 2D maps revealed in detail the spatial distribution of important parameters such as $kT$, $P$, and $S$, which confirmed that we are witnessing shock fronts,
induced by the merger of the subsystems. The high temperature shock regions shown in the 2D maps are spatially correlated  with higher entropy values, while our pressure maps showed an almost spherically symmetric region centered in between the clusters.

% simulations
Using tailored hydrodynamical $N$-body simulations, we obtained models that recover some of the observed features of the merging clusters. The main constraints were their virial masses, their subcluster separations and the shock temperatures. The challenge of simulating such collisions resides in the difficulty of simultaneously matching all the available constraints. The simulations presented here succeed in quantitatively reproducing the global features within an acceptable agreement.

% results G292
With our best model for G292, we find that this cluster merger is well represented by a frontal collision seen at an inclination of $18^{\circ}$. The observations are consistent with a simulated scenario where the two clusters are incoming for their first approach. The best instant (when the separation and temperatures are well matched) occurs 150\,Myr before the central passage. We may conclude that, in this scenario, we would be witnessing the G292 shortly before the encounter.

% results G036
For G036, our best model indicates that the collision is not frontal: it starts with a considerable impact parameter. At the moment of pericentric passage, the distance between the simulated clusters would be 146\,kpc. This may seem small, but even such short distances are known to induce considerable assymetries in the morphology \citep{Machado2015}. At the best instant, the plane of the orbit needs to be inclined by $50^{\circ}$ with respect to the plane of the sky. In this way, the separation and the projected temperature are quantitatively matched. This moment takes place only 50\,Myr before pericentric passage. Again, an accuracy of tens of Myr is surely not within the reach of this modelling. Still, we may conclude that the observational data for G036 is consistent with a scenario where we would be essentially witnessing the moment of pericentric passage, or a least very short time before it.

% mach
The simulated Mach numbers are roughly in the same range as the observational estimates. This results from the fact that they were both measured from the amplitudes of the temperature jumps across the shock fronts, and the best instants of the simulation had been chosen (among other criteria) such as to give the best possible quantitative correspondence with the observational temperature maps. The simulated Mach numbers are found to be slightly overestimated in both cases. We interpret the observational measurement to be a lower limit, since our best models suggest inclinations between the orbital planes and the plane of the sky. The true Mach numbers are generally found to be larger \citep[e.g. by $\sim30$ per cent in the case of A3376;][]{Machado2013}.

% caveats
Dedicated simulations such as these provide some useful insight into the history and dynamical state of the clusters. Although these reconstructions are plausible and physically well motivated, one should bear in mind that they are not unique. It is conceivable that alternative models could provide similarly acceptable agreements and thus other scenarios cannot be straightforwardly ruled out. Another shortcoming is that the simulations are highly idealized and fail to reproduce the morphology of the shock fronts in very fine details. Among other simplifications, they do not take into account any previous history during which the clusters might have interacted with additional substructures.

%\begin{itemize}
%\item G292 is a pre-merger system with a collision almost in the plane of the sky ($i=18^{\circ}$). These systems are 150 Myr

%\item With our high resolution  XMM-\textit{Newton} observations, two close (in projection) subclusters, G036N and G036S, were clearly resolved, and  spectroscopy could be performed. Spectral analysis reveals a higher temperature in the region between them compared to the symmetric regions away from their centers which could be due to shock heating during the merger. Our numerical simulation results confirmed that G036 is merging with the merger largely along the pane of the sky (50$^{\circ}$) and is 50\,Myr from the pericentric passage.

%\item  orientation effects may smear out the contrast across the pressure discontinuity and the X-ray surface brightness.
%\end{itemize}

%%%%%%%%%%%%%%%%%%%%%%%%%%%%%%%%%%%%%%%%%%%%%%%%%%%
\section*{Acknowledgements}
The authors thanks A.~Caproni for valuable suggestions. TFL acknowledges financial support from the Brazilian agencies FAPESP and CNPq through grants 2018/02626-8 and 303278/2015-3, respectively. REGM acknowledges support from the Brazilian agency CNPq
%\textit{Conselho Nacional de Desenvolvimento Cient\'ifico e Tecnol\'ogico} (CNPq) 
through grants 303426/2018-7 and 406908/2018-4. 
PAAL thanks the support of CNPq, grant 309398/2018-5.
This work  made use of the computing facilities of the Laboratory of Astroinformatics (IAG/USP, NAT/Unicsul), whose purchase was made possible by  FAPESP (grant 2009/54006-4) and the INCT-A. Simulations were carried out in part at the Centro de Computa\c c\~ao Cient\'ifica e Tecnol\'ogica (UTFPR). 

\bibliographystyle{mnras} 
\bibliography{refs}

%%%%%%%%%%%%%%%%% APPENDICES %%%%%%%%%%%%%%%%%%%%%
\appendix
\section{Analyzing the high-temperature region in G292 map}
\label{ApfontesG292}
The temperature map of G292 revealed a region  of high temperature in the west side of the shock region. Temperature reached almost 18 keV and is one of the highest values for a shock region in the literature. To make sure that this high-temperature values were due to the
shock and were not related to any point source which has not been detected previously, we investigated this region in other wavelengths: radio (VLA-First 1.4 Ghz and NVSS), optical (DSS and SDSS), and infrared  (IRAS, 2MASS, and WISE ). 

\begin{figure*}
\includegraphics[width=1.\textwidth]{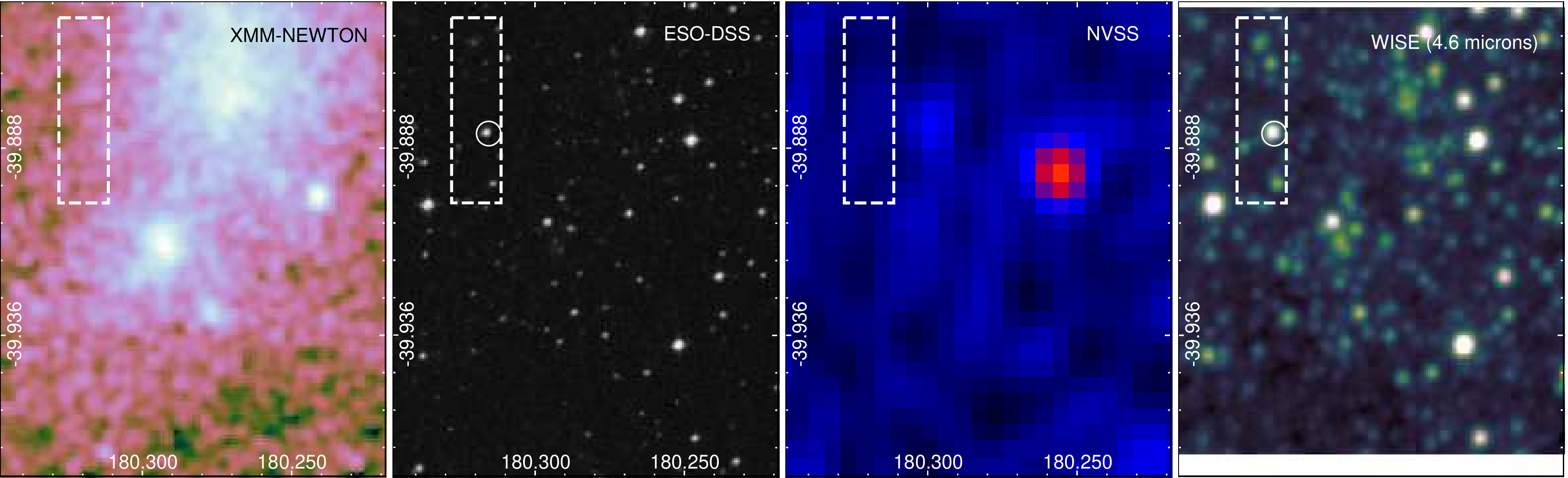} \\~\\
\includegraphics[width=0.287\textwidth,angle=-90]{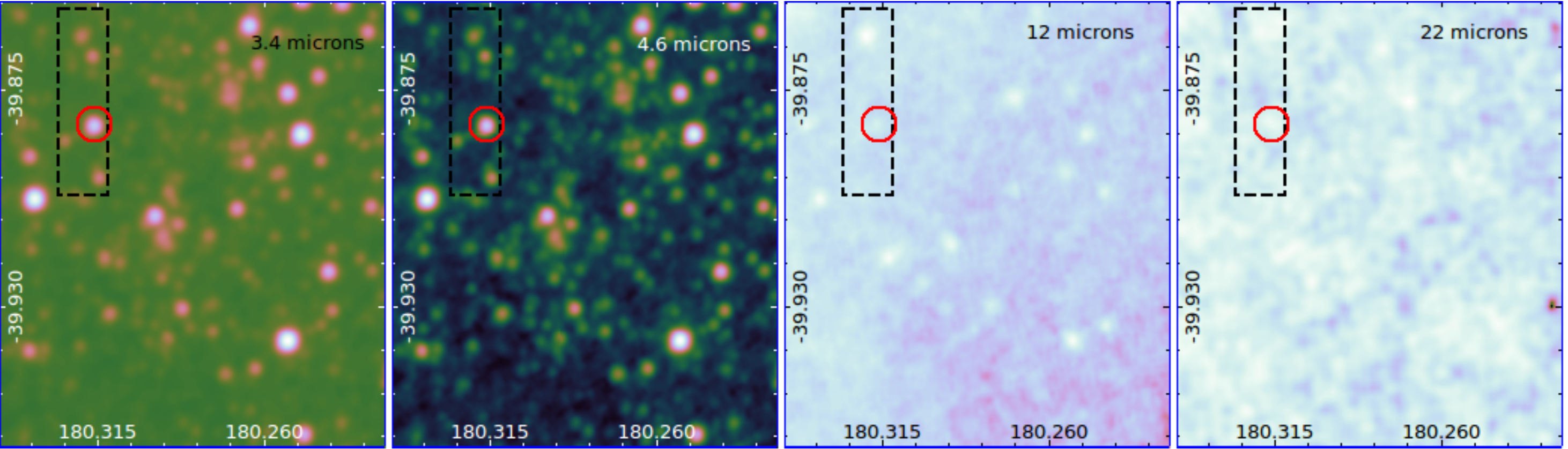}
\caption{\textit{Top panels}: XMM-\textit{Newton} X-ray, ESO-DSS optical, NVSS radio and WISE 4.6$\mu$m IR images with the white-dashed rectangle overlaid indicating the region of high temperature in the 2D map.\textit{Lower panel}: WISE images in 3.4$\mu$m, 4.6$\mu$m, 12$\mu$m, and 22$\mu$m
from left to right.}
\label{fig:G292Fontes}
\end{figure*}

We found a source that appears in the optical and in all IR images from  WISE (3.4$\mu$m, 4.6$\mu$m, 12$\mu$m, and 22$\mu$m) as shown in Fig.~\ref{fig:G292Fontes}. 
Although an IR source should not contribute to the bremsstrahlung X-ray spectral fit, we excluded from our
analysis to reconstruct our 2D maps. The results are the same and the shock region appears in
the same range of temperature confirming that we are witnessing one of the hottest shocks in 
the literature.

%--------------------------------------------------

%--------------------------------------------------

%--------------------------------------------------
%\begin{figure*}
%\includegraphics[width=0.7\textwidth]{figs/FontesG292.pdf}
%\caption{Caption}
%\label{fig:G292Fontes}
%\end{figure*}   
%--------------------------------------------------

\section{Goodness of the fit}

\begin{figure*}
\includegraphics[width=0.36\textwidth]{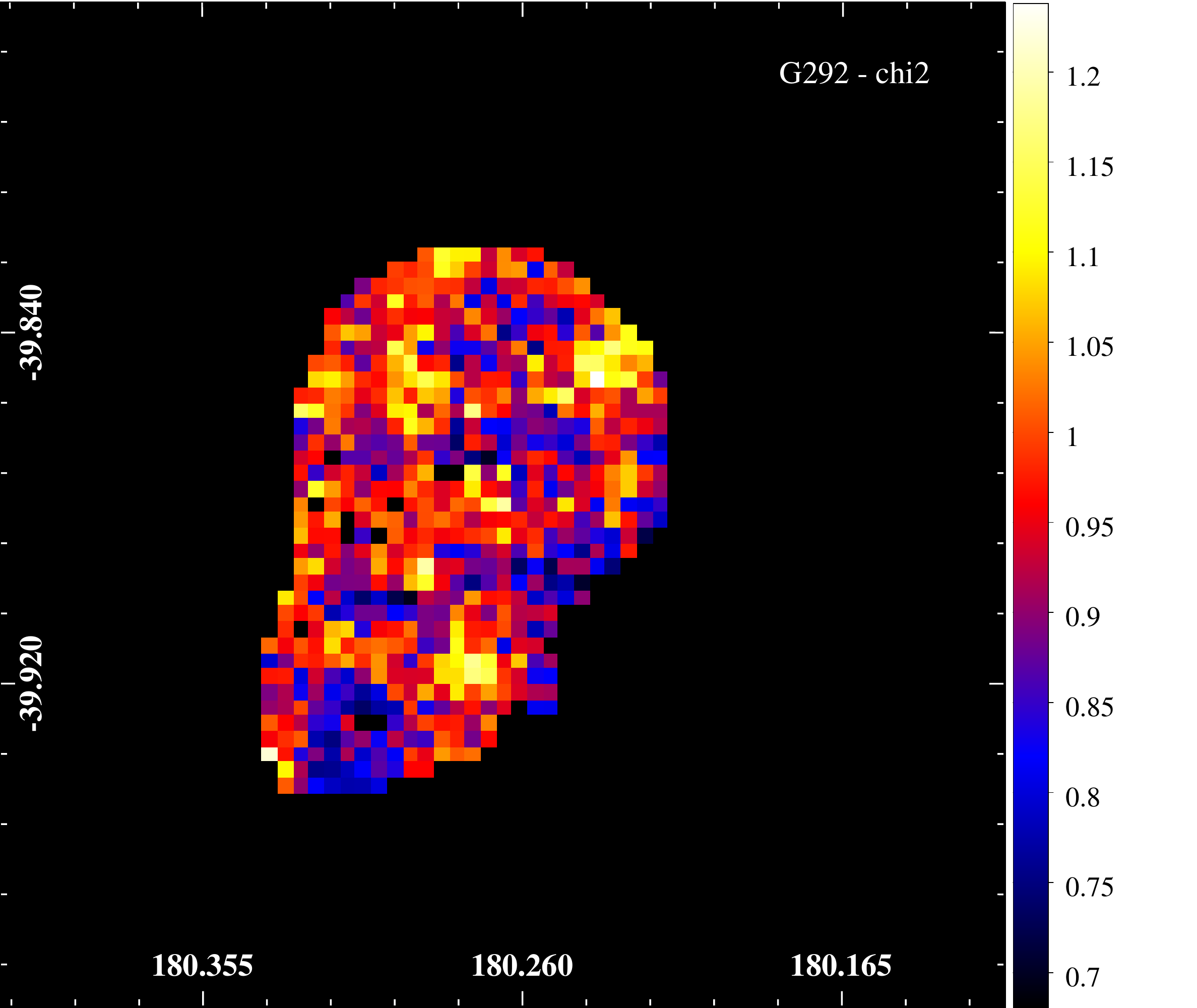}
\includegraphics[width=0.36\textwidth]{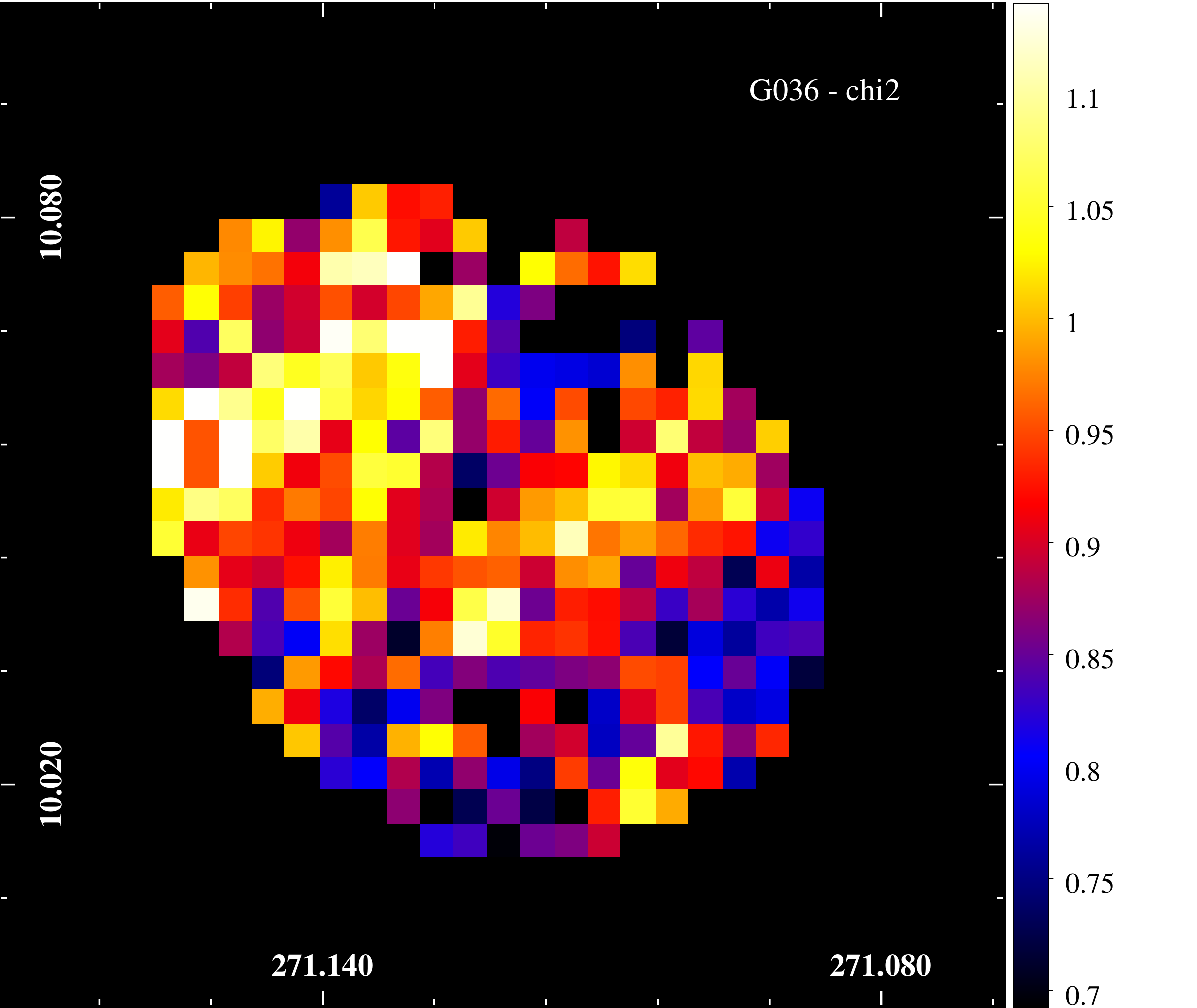}
\caption{Reduced $\chi^{2}$ Maps for G292 (left panel) and for G036 (right panel).}
\label{fig:MapasChi2}
\end{figure*}

To show the accuracy of each fit done to generate 2D maps, we show in Fig.~\ref{fig:MapasChi2} the Reduced $\chi^{2}$ Maps for G292 and G036. As it can be seen, all fits are in the 0.8-1.2 range. 

%--------------------------------------------------
%--------------------------------------------------

\bsp
\label{lastpage}
\end{document}